\documentclass[useAMS,usenatbib]{mn2e}

\usepackage{graphicx}
\usepackage{times}
\usepackage{natbib}
\usepackage{amsmath}

\newcommand{\gtrsim}{\ga}
\newcommand{\lesssim}{\la}

\def\zsun{{\rm Z_\odot}}
\def\msun{{\rm M_\odot}}
\def\msunh{{\rm M_\odot/{\it h}}}
\def\vb{$v_{\rm b}$}
\def\fnl{$f_{\rm NL}$}
\def\gnl{$g_{\rm NL}$}
\def\tnl{$\tau_{\rm NL}$}
\def\Mpch{{\rm Mpc/{\it h}}}
\def\kpch{{\rm kpc/{\it h}}}
\def\Omegat{{\Omega_{0,\rm tot}}}
\def\Omegab{{\Omega_{0,\rm b}}}
\def\Omegam{{\Omega_{0,\rm m}}}
\def\Omegal{{\Omega_{0,\rm \Lambda}}}

\title[Non-Gaussianities and first structures]{
The imprint of cosmological non-Gaussianities on primordial structure formation
}
\author[U.~Maio et al.]{
Umberto~Maio$^{1}$\thanks{E-mail: umaio@mpe.mpg.de},
Sadegh Khochfar$^{1}$
\\
${}^1$Max-Planck-Institut f\"ur extraterrestrische Physik,
Giessenbachstra{\ss}e 1,  D-85748 Garching b. M\"unchen, Germany
}

\begin{document}

\date{(draft)}
\pagerange{\pageref{firstpage}--\pageref{lastpage}}\pubyear{0}
\maketitle
\label{firstpage}

\begin{abstract}
We study via numerical N-body/SPH chemistry simulations the effects of primordial non-Gaussianities on the formation of the first stars and galaxies, and investigate the impact of supernova feedback in cosmologies with different \fnl.
Density distributions are biased to higher values, so star formation and the consequent feedback processes take place earlier in high-\fnl{} models and later in low-\fnl{} ones.
Mechanical feedback is responsible for shocking and evacuating the gas from star forming sites earlier in the highly non-Gaussian cases, because of the larger bias at high densities.
Chemical feedback translates into high-redshift metal filling factors that are larger by some orders of magnitude for larger \fnl, but that converge within one Gyr, for both population III and population II-I stellar regimes.
The efficient enrichment process, though, leads to metallicities $\gtrsim 10^{-2}\,\zsun$ by redshift $\sim 9$, almost independently from \fnl.
The impact of non-Gaussianities on the formation of dark-matter haloes at high redshift is directly reflected in the properties of the gas in these haloes, as models with larger \fnl{} show more concentrated gas profiles at early times.
Non-Gaussian signatures in the gas behaviour are lost after the first feedback takes place and introduces a significant degree of turbulence and chaotic motions.
Despite this, our results support the idea that non-Gaussianities can be imprinted in the gaseous and stellar features of primordial structures in the high-redshift Universe.

\end{abstract}



\section{Introduction}\label{Sect:introduction}


Among the several outstanding problems of modern Astrophysics and Cosmology \cite[][]{BarkanaLoeb2001,CiardiFerrara2005,BrommYoshida2011}, the status of the early Universe is certainly a very debated one.
The formation of primordial structures is supposed to be strongly influenced by the cosmological initial conditions they originated from, and it is widely accepted that all the visible objects derive from the growth of matter perturbations \cite[e.g.][]{GunnGott1972,WhiteRees1978,Peebles1993,Peacock1999,ColesLucchin2002,PR2003}, that developed shortly after the Big Bang, during the era of inflation.
These perturbations have grown in time in an expanding Universe and they have assembled into the galaxies, galaxy groups, and galaxy clusters observed today.
\\
According to recent determinations of the cosmological parameters \cite[e.g.][]{Komatsu2011}, the Universe is composed by $\sim 30\%$ of matter and for the remaining $\sim 70\%$ of an unknown term attributed to the so-called cosmological constant, $\Lambda$, or dark energy \cite[see also e.g.][for N-body/SPH chemistry simulations in dark-energy cosmologies and the effects on baryonic structure evolution]{Maio2006}.
More precisely, the present matter contributions to the cosmic density are \cite[][]{Komatsu2011}
$\Omegam = 0.272$,
$\Omegal = 0.728$,
$\Omegab = 0.044$,
for matter, cosmological constant, and baryons, respectively.
The cosmic equation of state parameter is consistent with $w=-1$,
the observed spectral index of primordial fluctuations is $n=0.96$, and 
the power spectrum normalization is given by a mass variance within $8~\rm\Mpch$-sphere $\sigma_8=0.8$.
\\
Structure formation depends strongly on the initial density fluctuations imprinted on the primordial matter distribution \cite[][]{PressSchechter1974,ShethTormen1999}.
The standard assumption on the distribution of density fluctuations in the Universe is based on an episode of dramatic size growth of the universe, roughly $10^{-37}$~s after the Big Bang, during which the seeds of the present-day structures formed \cite[][]{Starobinsky1980,Guth1981,Linde1990}.
These models of inflation predict in general that the overdensity $\delta$ is a Gaussian random variable with variance fully determined by the underlying power-spectrum \cite[e.g.][and references therein]{Komatsu2010,Casaponsa2011tmp,Curto2011arXiv,Bruni2011arXiv}.
The general consensus on the Gaussianity derives mainly from the central limit theorem.
However, observational determinations \cite[see Table in][]{MaioIannuzzi2011} show evidence for deviations from Gaussianities that leave room for non-Gaussian investigations, as well \cite[][]{Peebles1983,Viel2009,DesjacquesSeljak2010,LoVerde2011arXiv,Desjacques2011arXiv,DAmico2011,Hamaus2011arXiv}.
The effects of non-Gaussianities are expected to play a role mostly for the high-sigma density fluctuations \cite[e.g.][]{Grinstein1986,Koyama1999,Zaldarriaga2000,Wagner2010,LoVerde2011arXiv}, and, thus, very early structures should result somehow affected by them \cite[][]{Maio2011cqg}.
In particular, due to the sensitivity of the gas cooling capabilities to the underlying matter densities, the initially skewned non-Gaussian features could be reflected by the earlier collapse of molecular gas, and theoretically influence the formation epoch of the first stars and galaxies, as pointed out via numerical N-body/SPH chemistry simulations by \cite{MaioIannuzzi2011}.
Beyond the formation redshift, also the consequent feedback mechanisms could play an important role in ejecting material from the star forming regions or in polluting the surrounding environment.
Because of the lack of relevant studies dealing with non-Gaussianities and feedback mechanisms, these are still open questions which we will discuss and investigate throughout this work.
\\
In particular, it is possible to distinguish into mechanical, chemical, and radiative feedback \cite[for an extensive review of the many possible mechanisms, see e.g.][and references therein]{CiardiFerrara2005}.
The first class includes all the different phenomena related to mass or energy deposition into the cosmic gas from star formation and stellar evolution (i.e. shocks, blowout, blow-away, etc.); the second one comprises essentially the effects of chemical changes in the gas composition (i.e. metal enrichment and consequent stellar population transitions); and the third one covers the aspects linked to radiation emitted by cosmic sources (i.e. photoionization or photodissociation of molecules and atoms, gas photoheating, cosmic reionization, etc.).\\
We will mainly consider mechanical and chemical feedback from first structures, both from population III (popIII) and from population II-I (popII) stars.
The transition between these two regimes is determined by a critical metallicity, $Z_{crit}$ which, according to different authors \cite[][]{Schneider_et_al_2002,Schneider_et_al_2006,Bromm_Loeb_2003}, is estimated to be around $\sim 10^{-6}-10^{-3}\,Z_\odot$.
It has also been previously investigated in details \cite[e.g.][]{Tornatore2007,Maio2010, Maio2011b} with the help of numerical simulations following the chemical evolution of the cosmic gas and metal pollution from stars.
The substantial distinction between the popIII and the popII regime is the stellar initial mass function. While in the latter case, it is known to be Salpeter-like, in the former case it is unknown, even if expected to be top-heavy, for the incapability of pristine gas to cool down to very low temperatures, in presence of a high CMB floor (a few hundreds K at $z\simeq 10$).
However, fragmentation of primordial gas clouds and formation of popIII star with masses below $\sim 10^2\,\rm M_\odot$ can still be possible \cite[as shown by e.g.][]{Yoshida2006, Yoshida_et_al_2007,CampbellLattanzio2008,SudaFujimoto2010}.
The impacts of different assumptions on the primordial IMF, yields, and supernova ranges have already been largely studied by \cite{Maio2010}, thus we will not go into the details here and will simply assume a top-heavy IMF.
Stars with masses larger than $\sim 100\,\rm M_\odot$ are short-lived (up to $\sim 10^6\,\rm yr$) and the ones in the range [160, 240]~M$_\odot$ die as pair-instability SN (PISN) and are able to pollute the surrounding medium by ejecting large amounts of metals.
\\
In the present work, we will focus on the impacts of mechanical and chemical feedback in the primordial Universe and address their role in presence of non-Gaussian initial conditions.
The paper is structured as follows: after presenting in Sect.~\ref{Sect:simulations} the simulations used, in Sect.~\ref{Sect:results} we will discuss the main results related to the cosmological effects of mechanical feedback (Sect.~\ref{Sect:mechanical_feedback}) and chemical feedback (Sect.~\ref{Sect:chemical_feedback}), and we will also show the consequences for single haloes over cosmic time (Sect.~\ref{Sect:haloes}) and the implications of primordial streaming motions (Sect.~\ref{Sect:vbulk}).
In Sect.~\ref{Sect:discussion} we will summarize our findings and conclude.


\section{Simulations}\label{Sect:simulations}


The simulations considered here were firstly described in \cite{MaioIannuzzi2011}, who performed a large set of runs with different box sizes and resolutions.
Since we want to study in detail the joint non-linear effects of feedback mechanisms and primordial non-Gaussianities, we will focus on the high-resolution simulations, having a box size of $0.5\,\rm \Mpch$, and an initial gas particle mass of $\sim 40\,\msunh$.
Local non-Gaussianities are included by adding second-order perturbations to the Bardeen gauge-invariant potential \cite[e.g.][]{Salopek1990}:
\begin{equation}\label{eq:nong}
\Phi = \Phi_{\rm L} + f_{\rm NL} \left[ \Phi_{\rm L}^2 - <\Phi_{\rm L}^2> \right],
\end{equation}
with $\Phi_{\rm L}$ the {\it linear} Gaussian part, and \fnl the dimensionless coupling constant controlling the magnitude of the deviations from Gaussianity.
Observational constraints on \fnl{} suggest values between $\sim 0-100$ \cite[a complete table of observational determinations is given in][]{MaioIannuzzi2011}, so we will focus on the cases \fnl=0, \fnl=100, and also on \fnl=1000 for sake of comparison.
\\
The simulations were performed by using a modified version of the parallel tree/SPH Gadget-2 code \cite[][]{Springel2005}, which included gravity and hydrodynamics,
with radiative gas cooling both from molecules and atomic transitions \cite[according to][]{Maio2007},
multi-phase model \cite[][]{Springel2003} for star formation,
UV background radiation \cite[][]{HaardtMadau1996},
wind feedback \cite[][]{Springel2003,Aguirre_et_al_2001},
chemical network for e$^-$, H, H$^+$, H$^-$, He, He$^+$, He$^{++}$, H$_2$, H$_2^+$, D, D$^+$, HD, HeH$^+$ \cite[e.g.][ and references therein]{Yoshida2003,Maio2006,Maio2007,Maio2009,Maio2009PhDT,Maio2010}, 
and metal (C, O, Mg, S, Si, Fe) pollution from popIII and/or popII stellar generations, ruled by a critical metallicity threshold of $Z_{crit}=10^{-4}\,\zsun$ \cite[][]{Tornatore2007,Maio2010,Maio2011b}.
The cosmological parameters are fixed by assuming a concordance $\Lambda$CDM model with
matter density parameter $\Omega_{\rm 0,m}=0.3$,
cosmological density parameter  $\Omega_{\rm 0,\Lambda}=0.7$,
baryon density parameter $\Omega_{\rm 0,b}=0.04$,
expansion rate at the present of H$_0=70\,\rm km/s/Mpc$,
power spectrum normalization via mass variance within 8~Mpc/{\it h} radius sphere $\sigma_8=0.9$,
and spectral index $n=1$.
We consider a top-heavy stellar initial mass function (IMF) with mass range [100, 500]~$\msun$ for the population III regime \cite[different cases are discussed in][]{Maio2010,MaioIannuzzi2011,Maio2011cqg}, and a Salpeter IMF with mass range [0.1, 100]~$\msun$ for the population II-I regime.
\\
A friend-of-friend (FoF) algorithm with comoving linking length of 20 per cent the mean inter-particle separation, is applied at post-processing time to find the formed cosmic objects, with their dark, gaseous, and stellar components.


\section{Results}\label{Sect:results}
In the following we will present results related to the interplay of mechanical feedback and chemical feedback with non-Gaussianities (Sect.~\ref{Sect:mechanical_feedback} and Sect.~\ref{Sect:chemical_feedback}), and the implications for early cosmic structures (Sect.~\ref{Sect:haloes} and Sect.~\ref{Sect:vbulk}).


\subsection{Mechanical feedback and gaseous properties}\label{Sect:mechanical_feedback}
We begin our discussion by commenting on the thermodynamical properties of the cosmic gas and the feedback mechanisms related to stellar evolution and supernova explosions.


\subsubsection{Distributions}

\begin{figure*}
\centering
\includegraphics[width=0.33\textwidth]{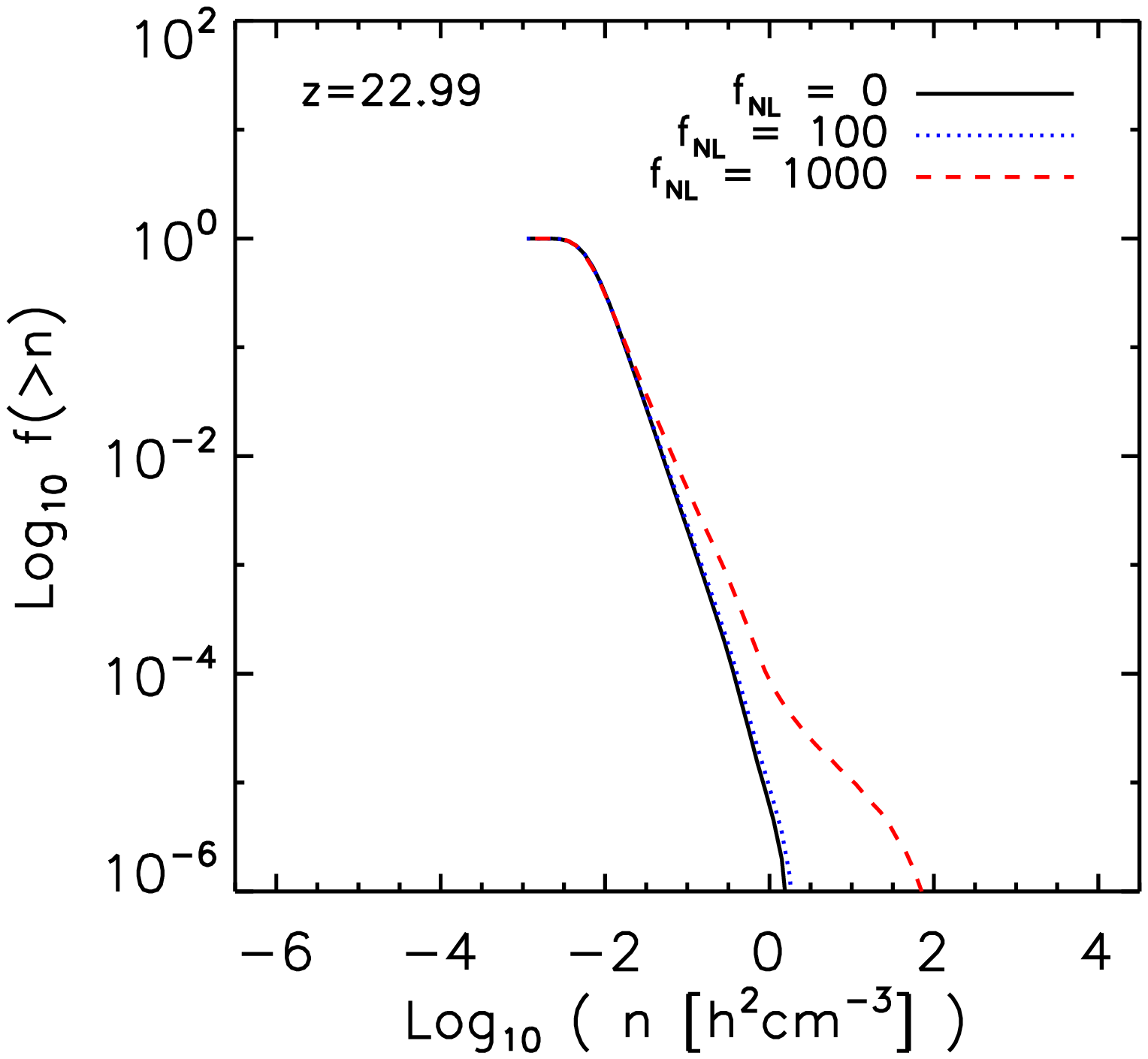}
\includegraphics[width=0.33\textwidth]{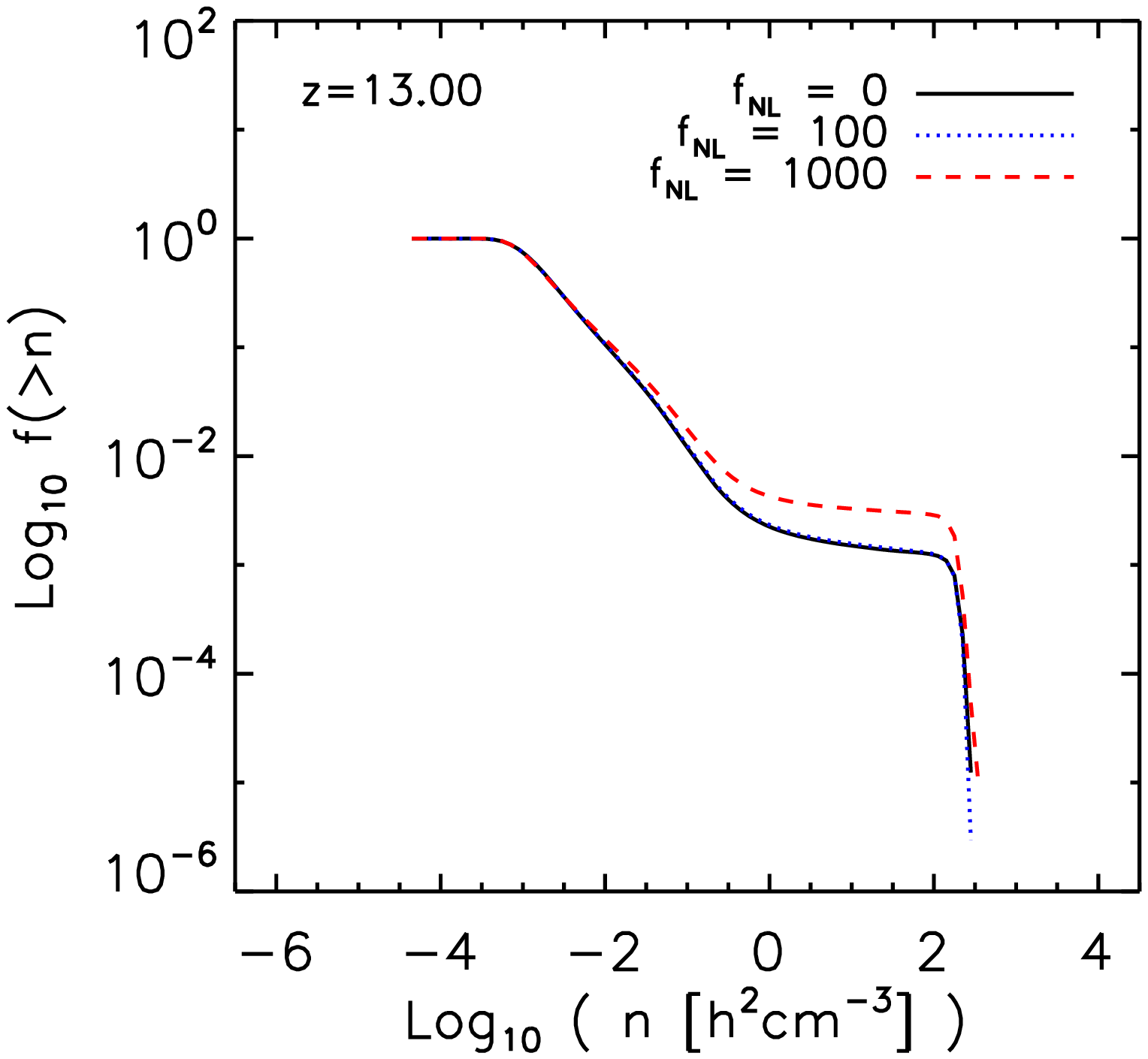}
\includegraphics[width=0.33\textwidth]{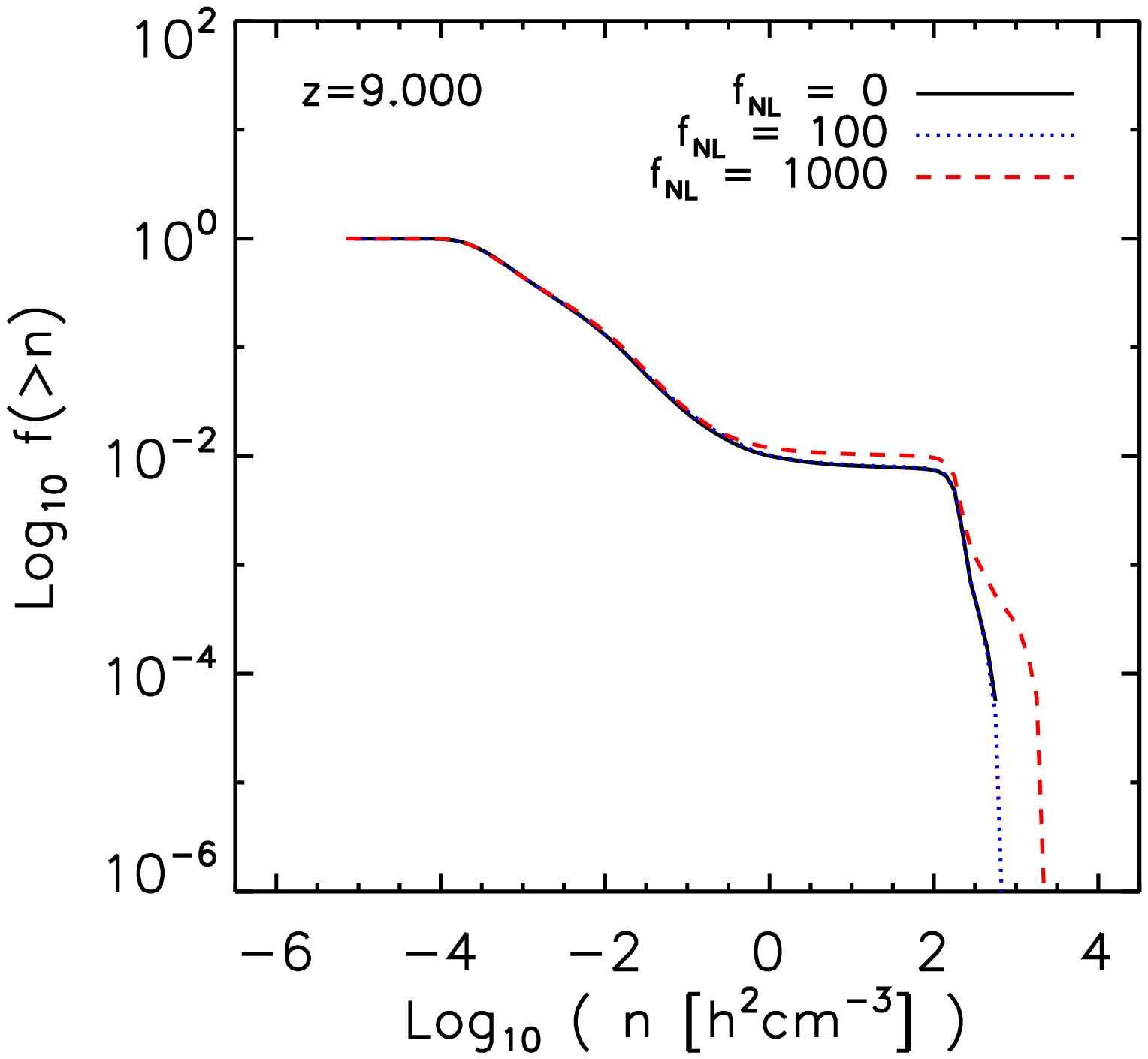}
\caption[Redshift comparison]{\small
Gas cumulative probability distributions, $f(>n)$, as a function of number density, $n$, at redshift, $z\simeq 23$ (left), $z\simeq 13$ (center), and $z\simeq 9$ (right), for \fnl=0 (solid lines), \fnl=100 (dotted lines), and \fnl=1000 (dashed lines), as indicated by the labels.
}
\label{fig:pdf05}
\end{figure*}

As already mentioned, non-Gaussianities play a role on the primordial matter distribution.
Therefore, we start our investigations by studying the gas mass distribution for the different \fnl{} scenarios.
We note that star formation sets in when the gas reaches a density of $\sim 10^2\, h^2\rm cm^{-3}$.
At this point, feedback mechanisms become active, too.
In Fig.~\ref{fig:pdf05}, we plot the cumulative mass fraction of gas having density larger than a given value, at redshift $z\simeq 23$, $z\simeq 13$, and $z\simeq 9$ for \fnl=0, \fnl=100, and \fnl=1000.
At $z\sim 23$ (left panel), most of the gas is still in low density environment, with only a small fraction of $\sim 10^{-3}$ undergoing first collapsing episodes, at number densities $n \gtrsim 0.1-1\,h^2\rm cm^{-3}$.
The \fnl=0 and \fnl=100 models are almost identical, while the \fnl=1000 case shows a larger tail at the high-density end, demonstrating that H$_2$ and HD molecules have been more efficiently formed and the molecular gas content has already reached fractions of $\gtrsim 10^{-2}-10^{-1}$.
This allows the gas to cool and condense more rapidly and to reach $n\sim 10^2 \, h^2\rm cm^{-3}$, while in the other models densities of $n\sim 1 \, h^2\rm cm^{-3}$ are barely achieved.
At later times, when $z\simeq 13$ (central panel), the density range is equally covered for all the \fnl{}, and the contribution from clumped regions increases of a few orders of magnitude.
The \fnl=1000 cosmology preserves some signatures of the primordial distribution, and these are reflected in a factor of $\sim 2-3$ in the higher distribution for $n> 0.1 \, h^2\rm cm^{-3}$.
Finally, at $z\simeq 9$ (right panel), the behaviours converge with some per cent of the gas collapsing into the dense star forming regions and almost erasing any evidence of non-Gaussianity.
Residual contributions persist around $n\sim 10^3-10^4\, h^2\rm cm^{-3}$ for \fnl=1000, though.


\subsubsection{Global evolution}
\begin{figure}
\centering
\includegraphics[width=0.5\textwidth]{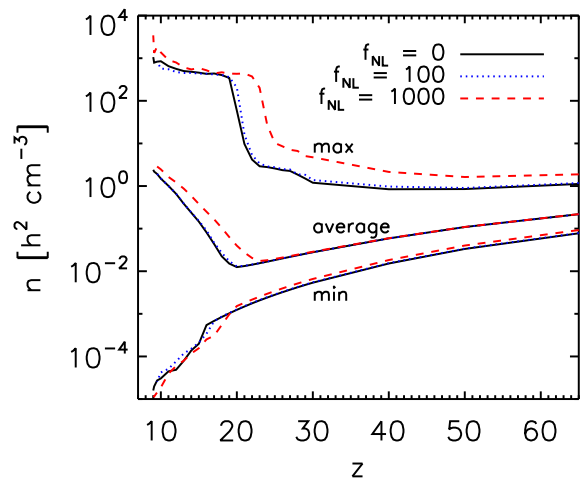}
\caption[Redshift comparison]{\small
Redshift evolution of the gas densities in the simulations for \fnl=0 (solid lines), \fnl=100 (dotted lines), and \fnl=1000 (dashed lines).
The data related to maximum, average and minimum values for each model are plotted from top to bottom, as indicated by the labels.
}
\label{fig:compare}
\end{figure}
In order to discuss more in depth mechanical feedback and its implications in Gaussian and non-Gaussian cosmologies, we study the global behaviours of gas density over cosmic time.
\\
In Fig.~\ref{fig:compare}, we plot the redshift evolution of gas densities, by focusing on the average density, the minimum density, and the maximum density reached in the runs.
\\
We note that the mean mass density, i.e. the ratio between the total mass in the box and the simulated volume does not give any significant information on the different models, since it evolves $\propto (1+z)^3$ independently from \fnl{}.
Thus, in order to see local signatures of non-Gaussianities on the gas behaviour and on its collapsing phases, we consider the average value of the densities of each gas particle, computed in the SPH fashion by smoothing the surrounding mass over the SPH kernel (in different words, the average we are showing is the average of the density PDF in each of the simulations).
In this way we can easily identify effects from the underlying \fnl{} on structure growth.
In fact, larger \fnl{} will enhance structure formation and hence hydrodynamical densities will increase more rapidly.\\
According to the figure, the average densities follow the cosmic average evolution fairly well and decrease smoothly down to redshift $z\sim 20$ (corresponding to $\sim 10^{-2}\,h^2\rm cm^{-3}$), when gas collapses and star formation episodes become effective, and boost the mean values by $\sim 2$ orders of magnitude.
This is well seen in the deviation from the decreasing trends from high to low redshift which signs the departure from the mean cosmic value and the onset of star formation.
In fact, at $z \lesssim 20$, when first highly non-linear structures form, densities increase locally of several orders of magnitude, and dominate over the typical values of the other particles.
Basically, the different averages for different \fnl{} at lower redshifts reflect the different non-linear behaviours of the very first collapsing objects.\\
The universes with \fnl=0 and \fnl=100 have very similar averages, whereas the model with \fnl=1000 shows earlier deviations because of the earlier collapse phases undergone by the gas.
The trend of the \fnl=1000 case is easily understood when looking at the maximum densities sampled by the simulations.
Both for \fnl=0 and \fnl=100 densities at early times ($z\gtrsim 20$) are around $\sim 1-10\,h^2\rm cm^{-3}$, instead, for \fnl=1000 they are systematically higher of a factor of a few and, thus, can grow faster, due to potentially enhanced cooling instabilities.
This is well visible during the first collapse, around $z\sim 20$, when the maximum density increases exponentially of $\sim 2$ orders of magnitude in $\sim 20$~Myr, and the discrepancies between \fnl=1000 and \fnl=100 or \fnl=0 are particularly evident.
The minimum densities do not show significant modifications with non-Gaussianities.
\\
In all the models, early evolution is characterized by a medium which is quite cold (at temperatures of a few hundreds Kelvin) and that is cooled predominantly by primordial molecules.
Shock heating of gas while it is falling into the dark-matter potential wells causes temperature increase up to $\sim 10^3-10^4\,\rm K$.
When the first star formation episodes take place (at $z\sim 20$) in the densest regions, stellar feedback, through PISN/SN explosions, rapidly brings the temperatures to $\sim 10^5-10^6\,\rm K$, several orders of magnitude hotter than the cooling medium.
Hot gas shocks the surrounding material and pushes the gas to lower-density regions, as well (it is the simultaneous ongoing gas collapse in different sites keeping high the maximum densities of Fig.~\ref{fig:compare}).
These sudden changes allow us to detect the earlier structure evolution mainly in the \fnl=1000 universe.
These stages are very short, though, and in some tens of Myr the different values in three models converge and the discrepancies fade away.


\subsection{Chemical feedback and metal enrichment}\label{Sect:chemical_feedback}
\begin{figure*}
\centering
\includegraphics[width=0.33\textwidth]{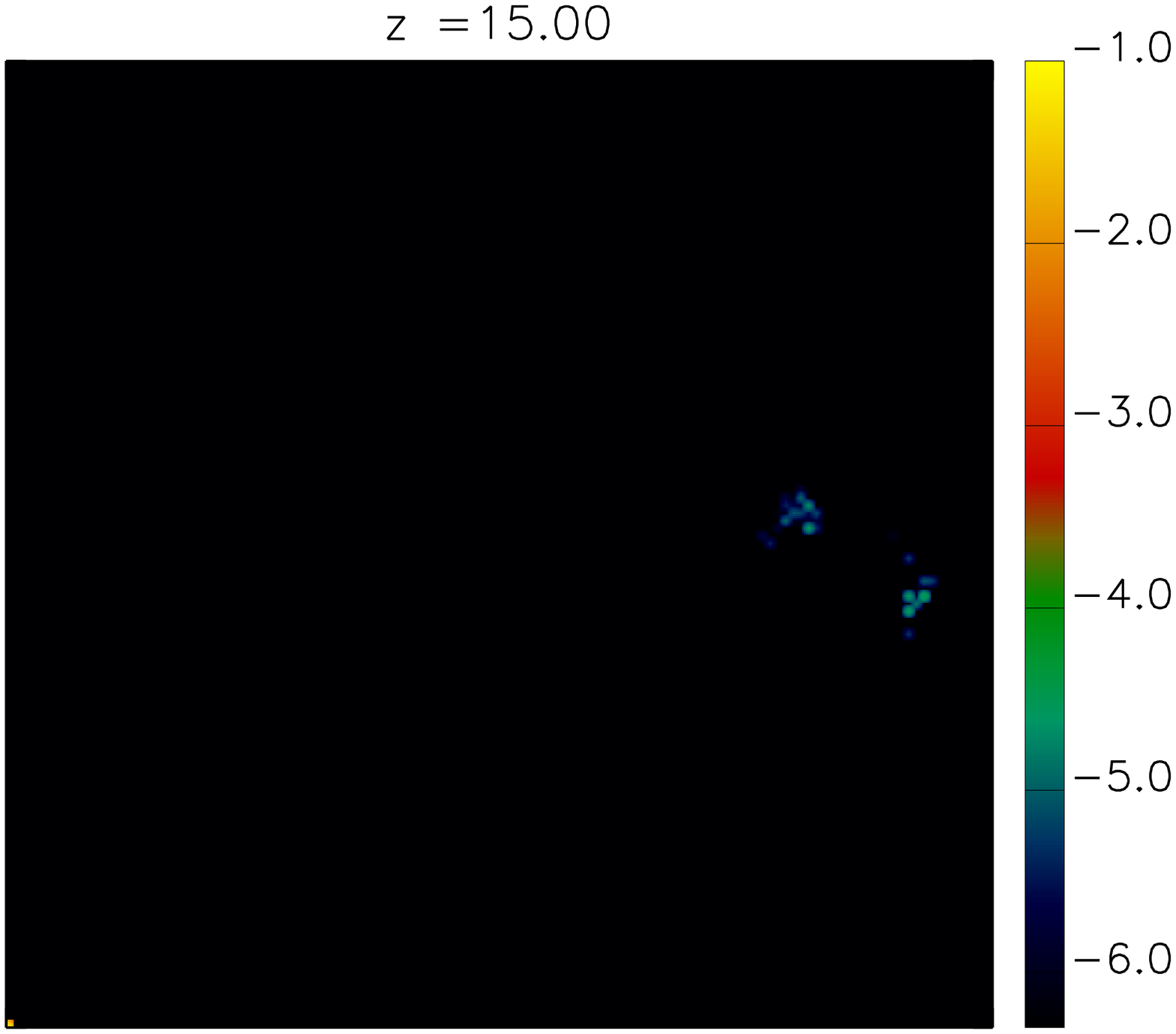}
\includegraphics[width=0.33\textwidth]{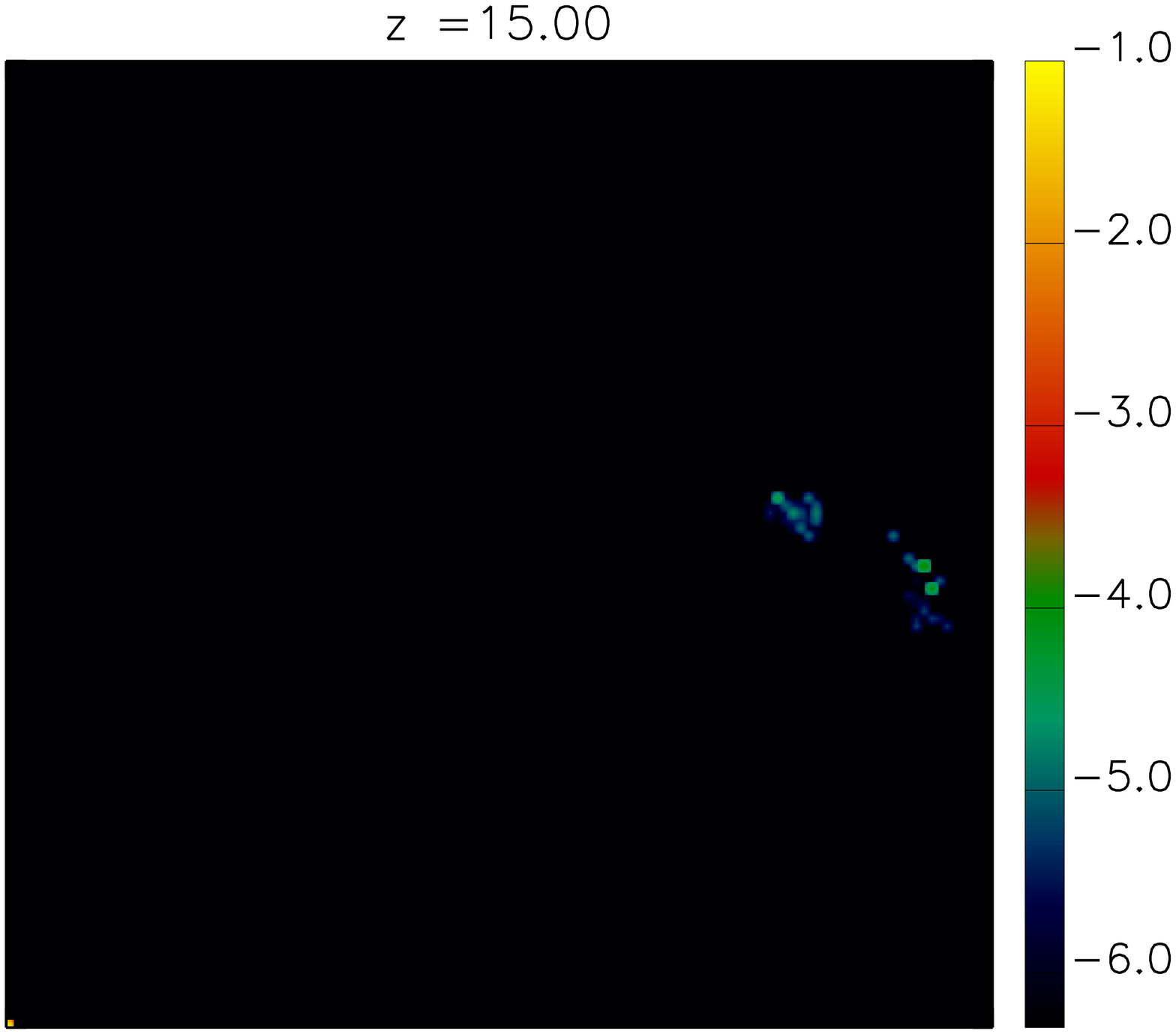}
\includegraphics[width=0.33\textwidth]{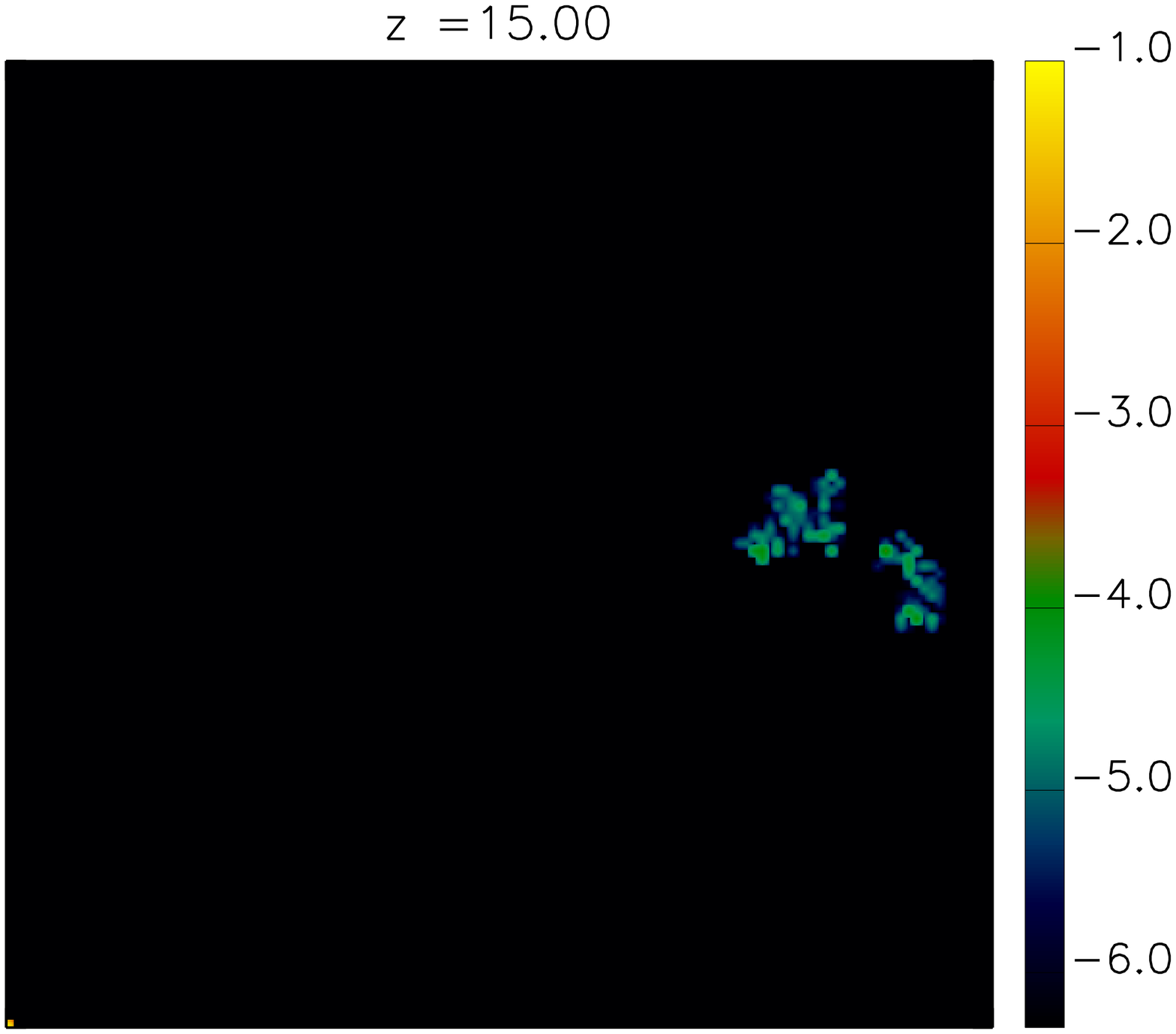}\\
\includegraphics[width=0.33\textwidth]{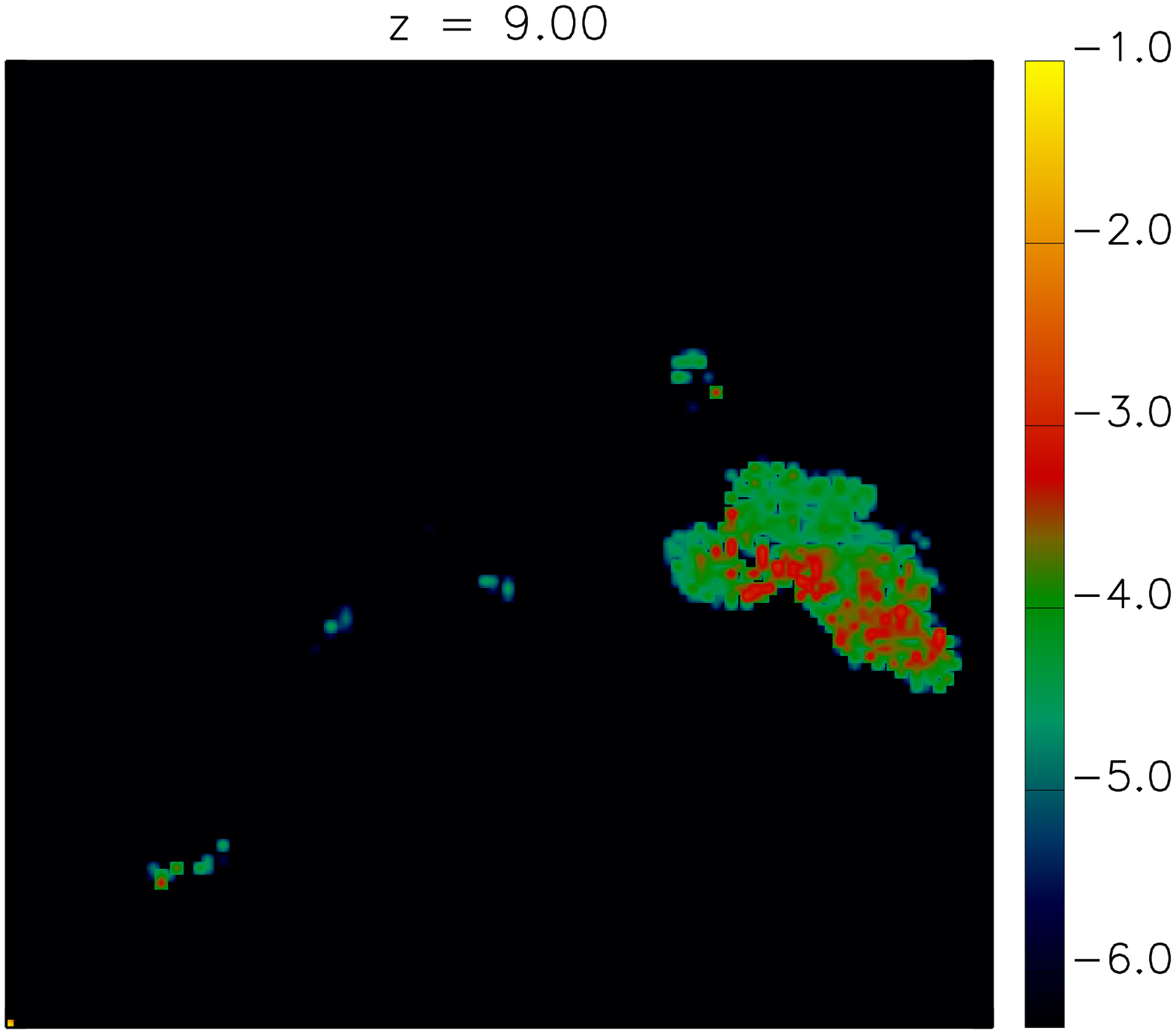}
\includegraphics[width=0.33\textwidth]{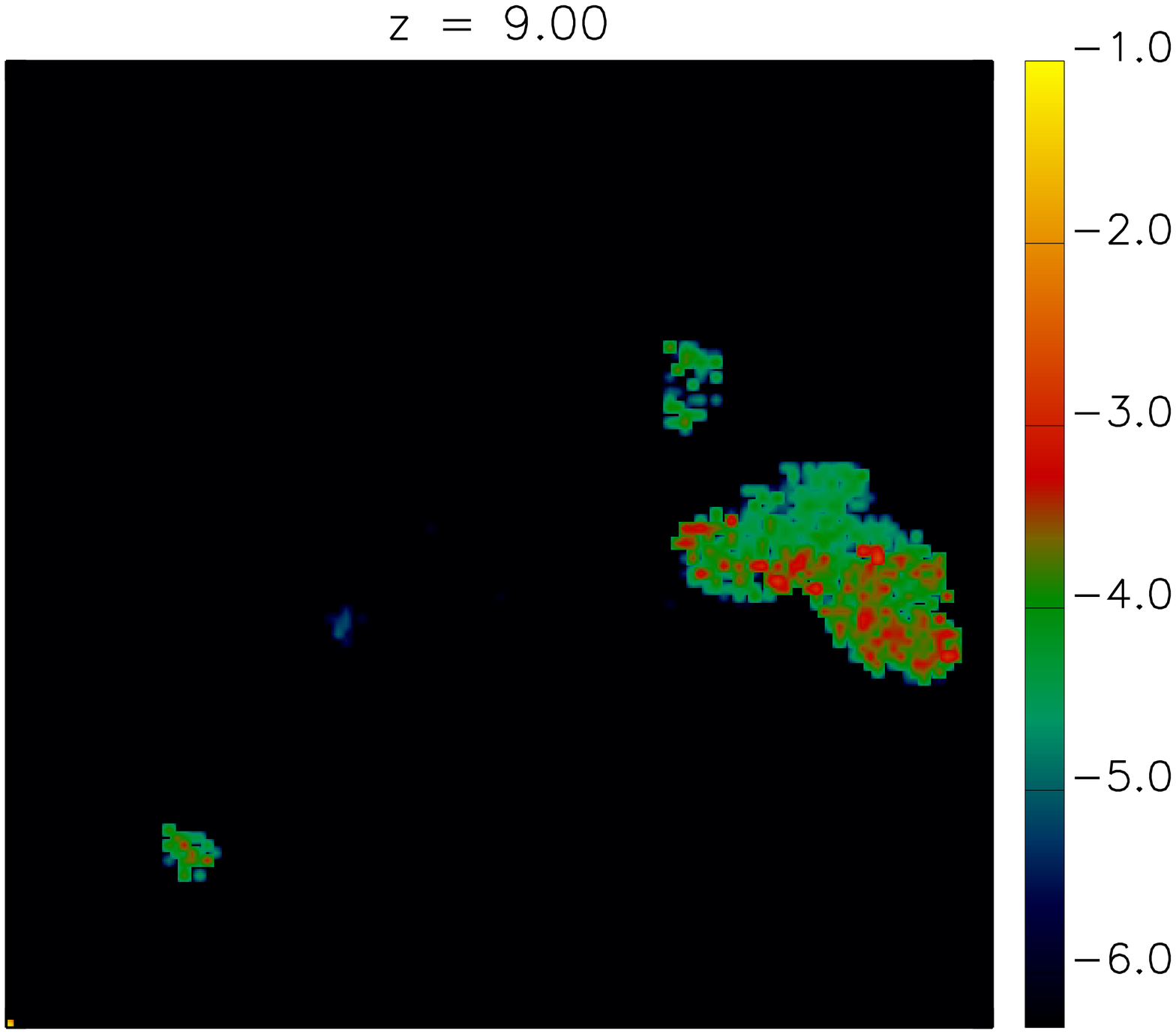}
\includegraphics[width=0.33\textwidth]{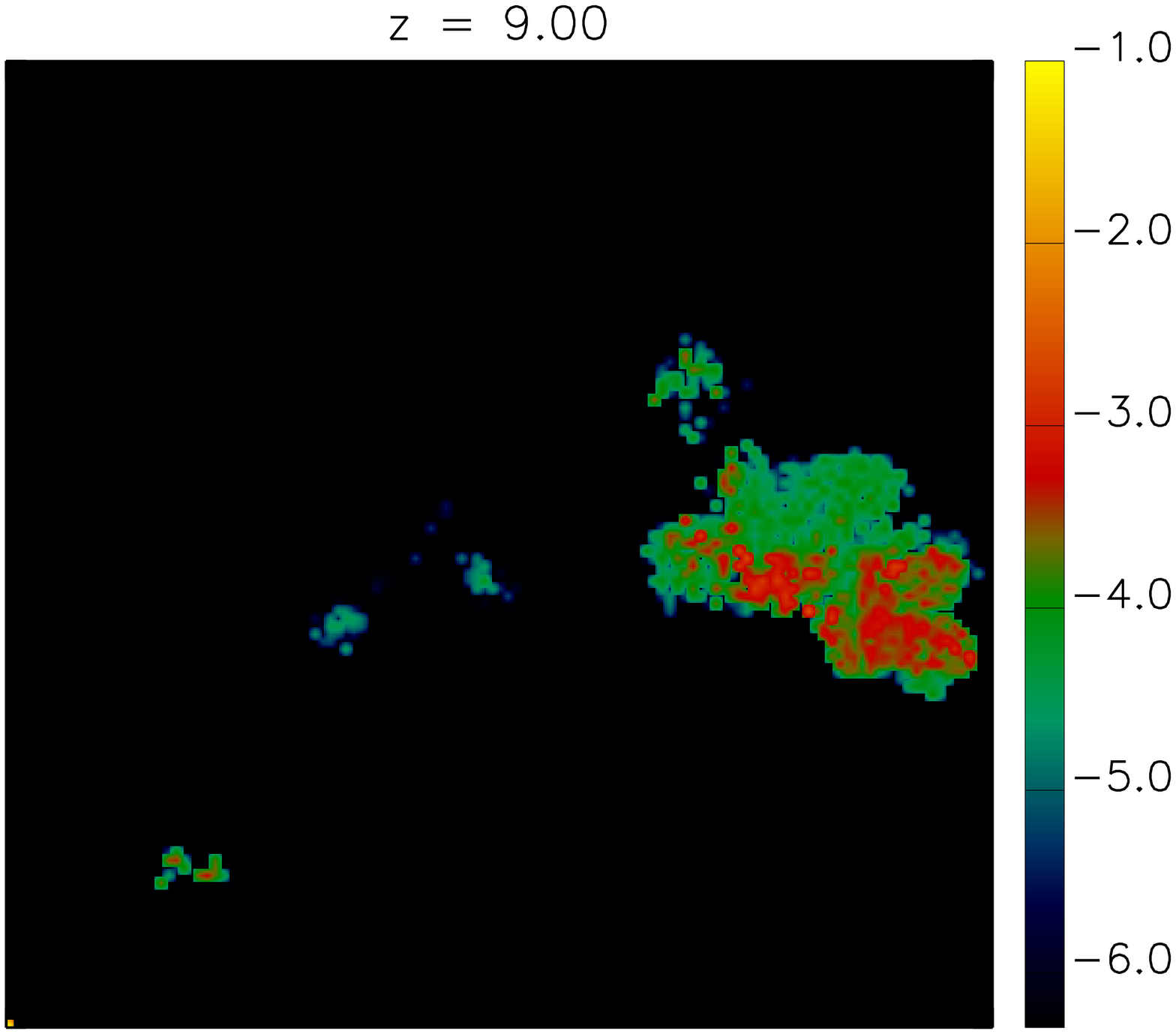}
\begin{tabular}{ccc}
\fbox{\fnl=0} & \hspace{0.33\textwidth}\fbox{\fnl=100}\hspace{0.33\textwidth} & \fbox{\fnl=1000} \\
\end{tabular}\\
\caption[Map evolution]{\small
Metallicity maps for \fnl=0 (left column), \fnl=100 (central column), and \fnl=1000 (right column), at $z=15$ (upper row) and $z=9$ (lower row).
The metal mass included in a slide thick $\sim 1/14$ the box size has been projected and smoothed according to the SPH kernel, on a grid of 128$\times$128 pixels, on the plane at height $z_0=250\,\kpch$.
}
\label{fig:maps}
\end{figure*}

At this point, we discuss the main results related to the chemical feedback and its effects, in particular metal pollution at early times.
A pictorial representation of the pollution events in the different boxes is given in Fig.~\ref{fig:maps}, where we display the metallicity maps for \fnl=0 (left), \fnl=100 (center), and \fnl=1000 (right), at $z=15$ (upper row) and $z=9$ (lower row).
These immediately show some differences among the various cases, mostly at $z=15$, when the first spreading events are taking place.
For all the cases, enrichment up to $Z/H \sim 10^{-4}$ is detected, but it is more limited for \fnl=0 and \fnl=100 than for \fnl=1000.
At redshift $z=9$, the amount of metals spread is comparable, but, as we will better quantify in the next sections, in the \fnl=1000 there is slightly more enrichment that results in a larger filling factor.


\subsubsection{Phase distributions}
\begin{figure*}
\centering
\begin{tabular}{ccc}
\fbox{\fnl=0} & \hspace{0.33\textwidth}\fbox{\fnl=100}\hspace{0.33\textwidth} & \fbox{\fnl=1000} \\
\end{tabular}\\
\includegraphics[width=0.33\textwidth]{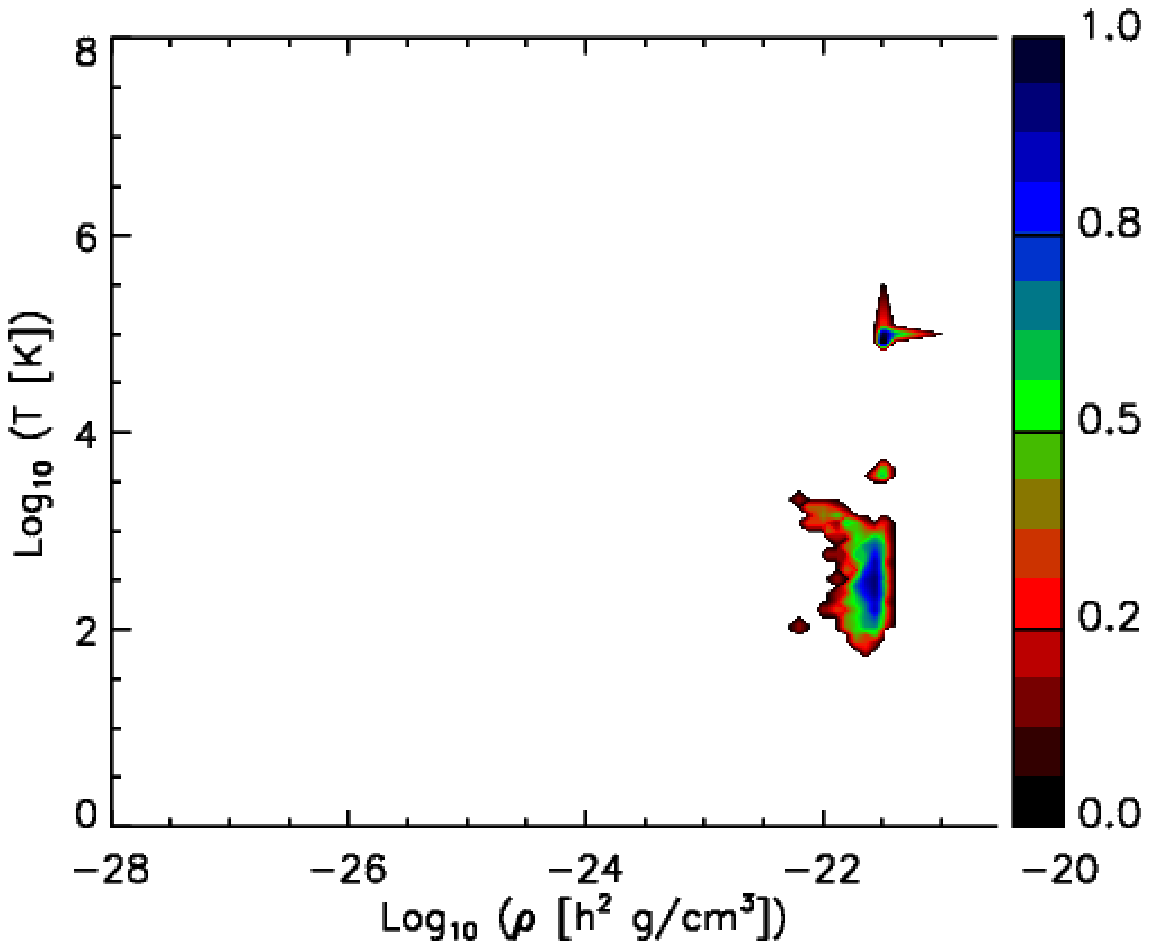}
\includegraphics[width=0.33\textwidth]{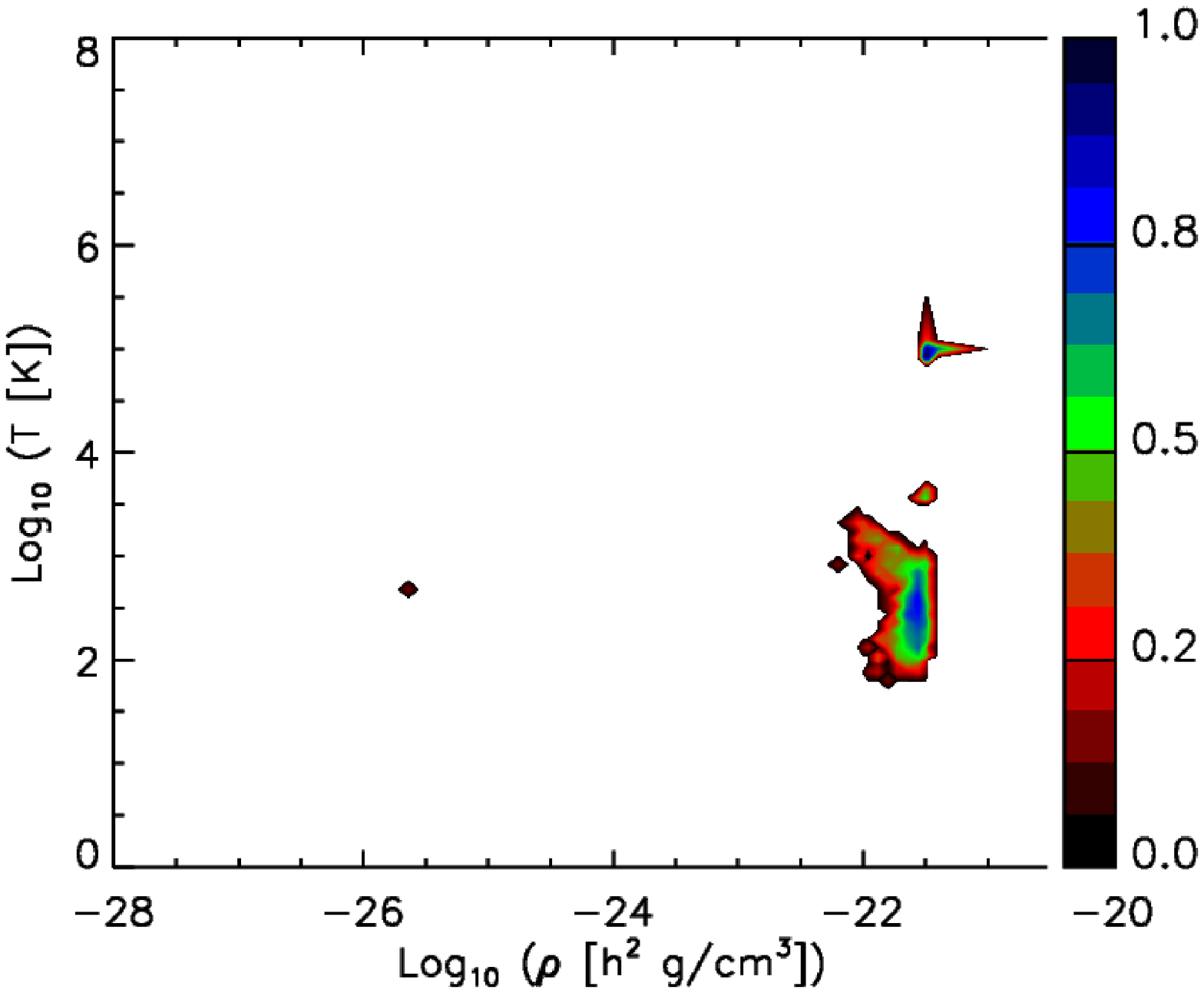}
\includegraphics[width=0.33\textwidth]{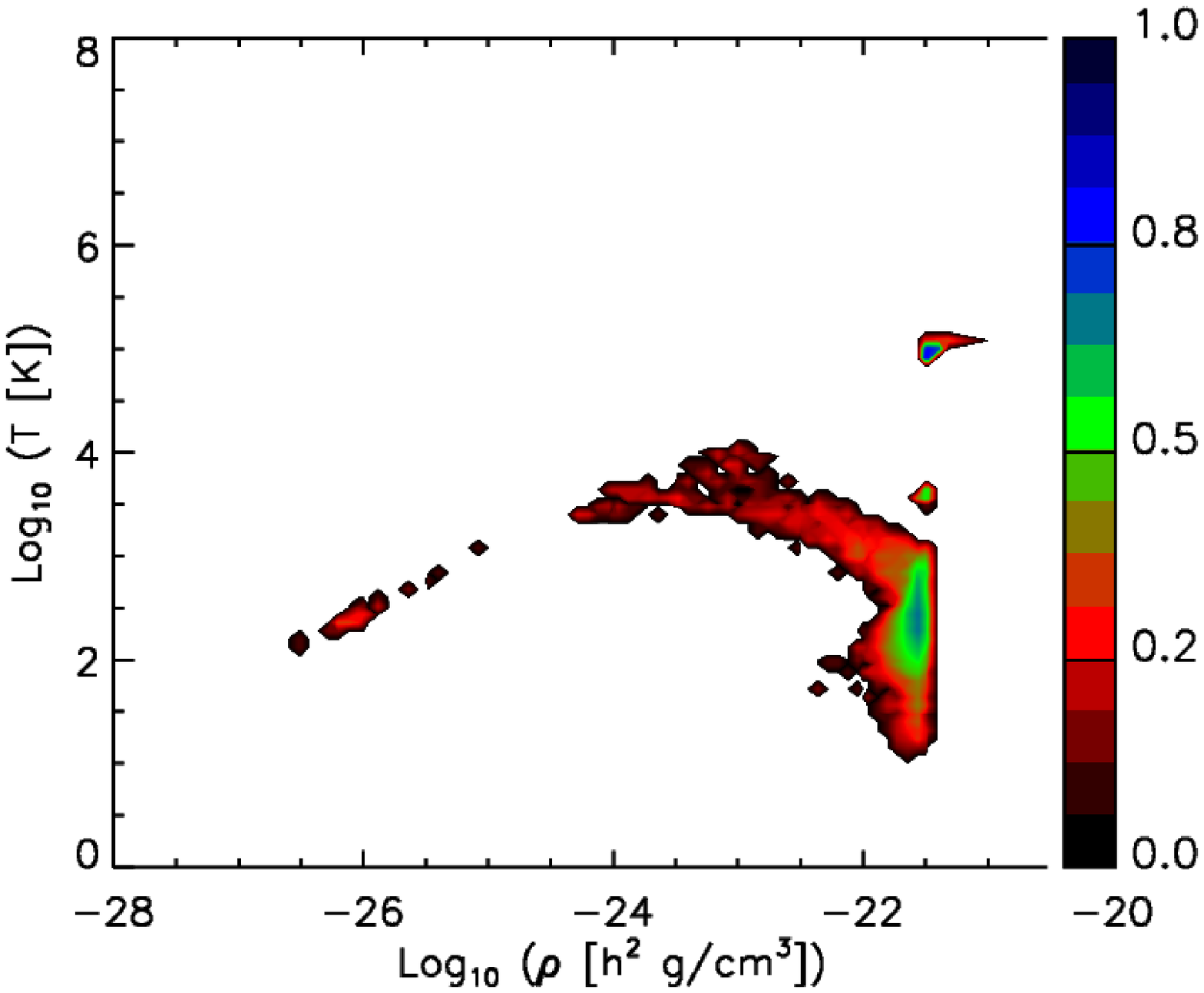}\\
\flushleft{$z=18.01$}\\
\includegraphics[width=0.33\textwidth]{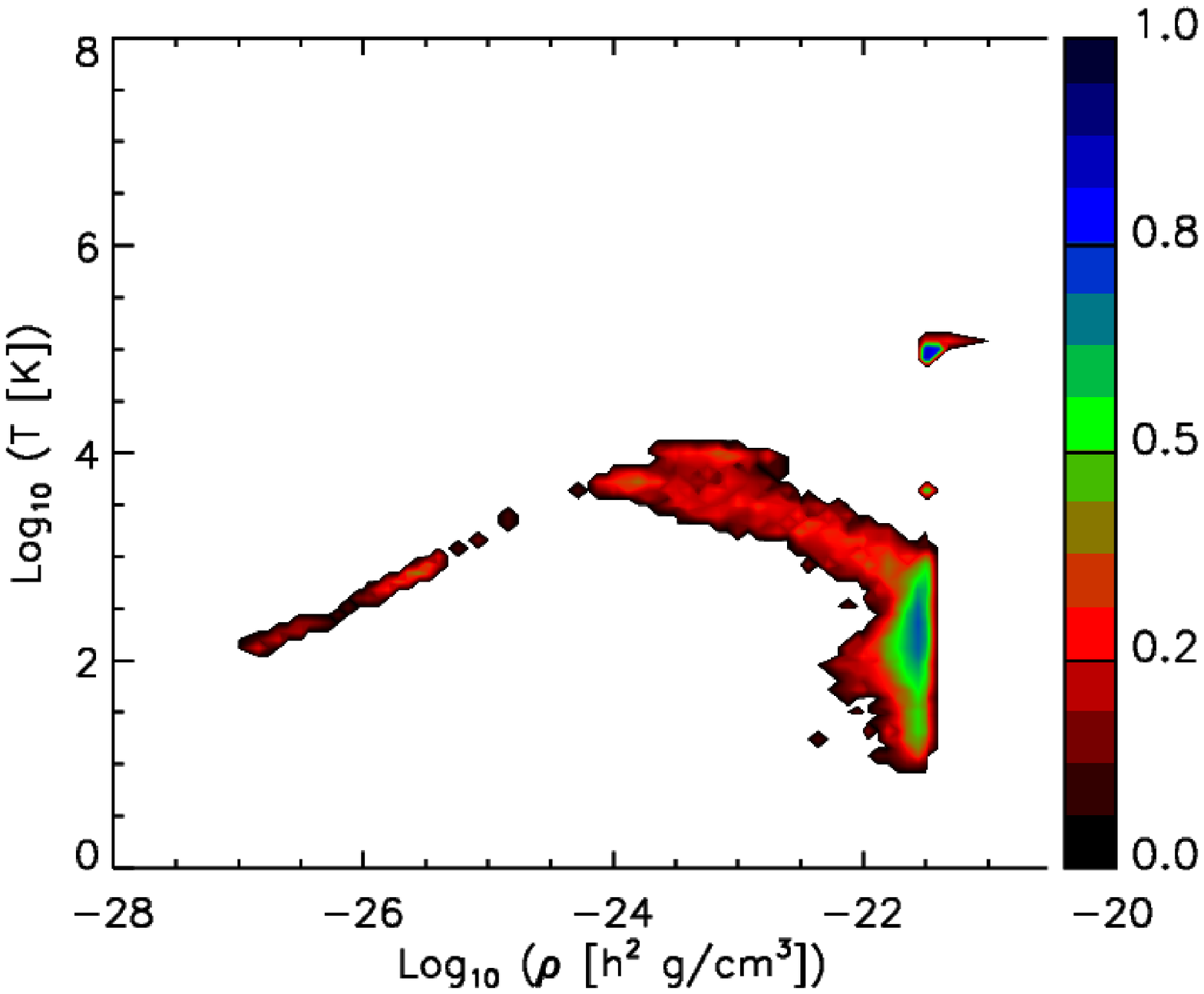}
\includegraphics[width=0.33\textwidth]{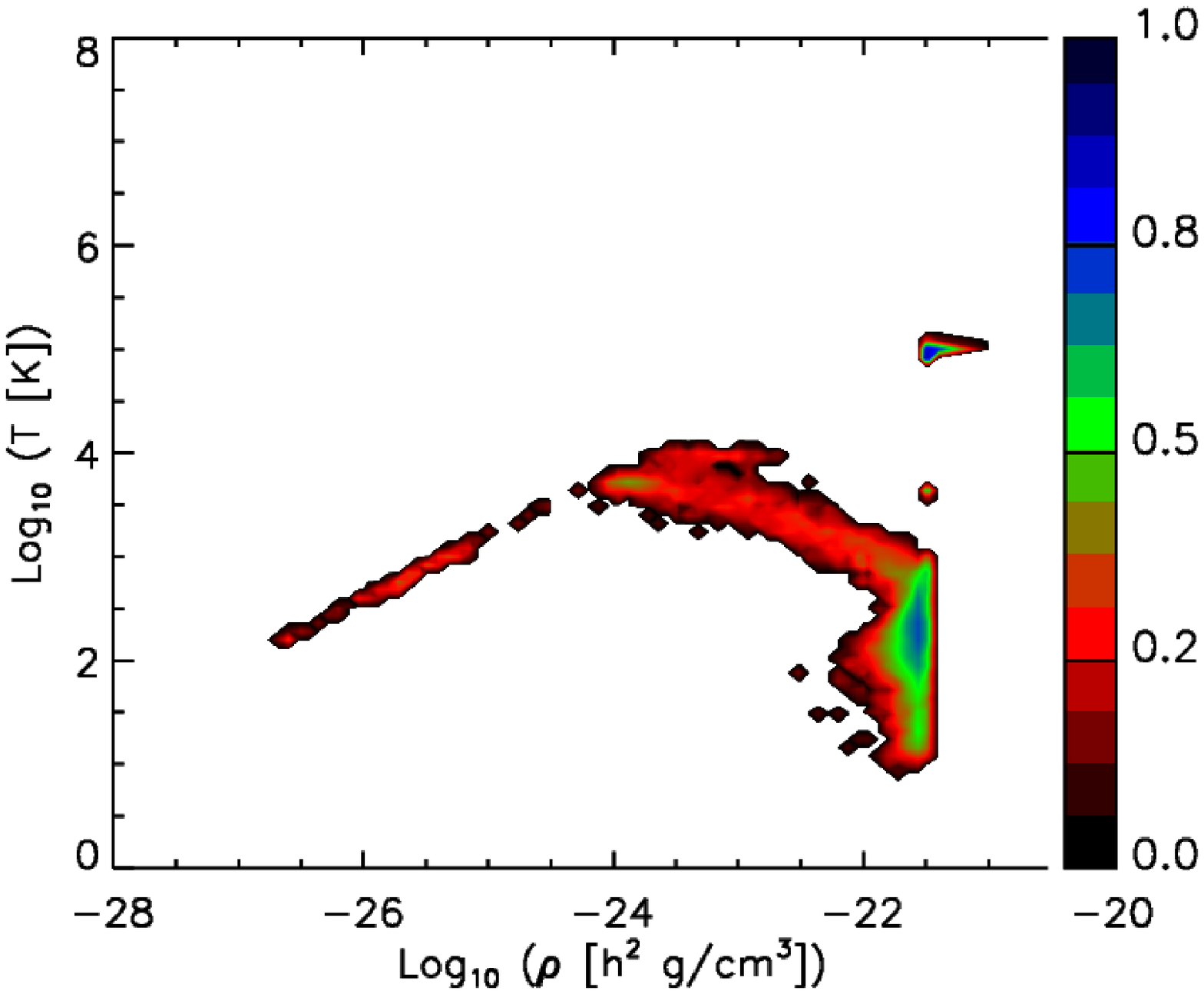}
\includegraphics[width=0.33\textwidth]{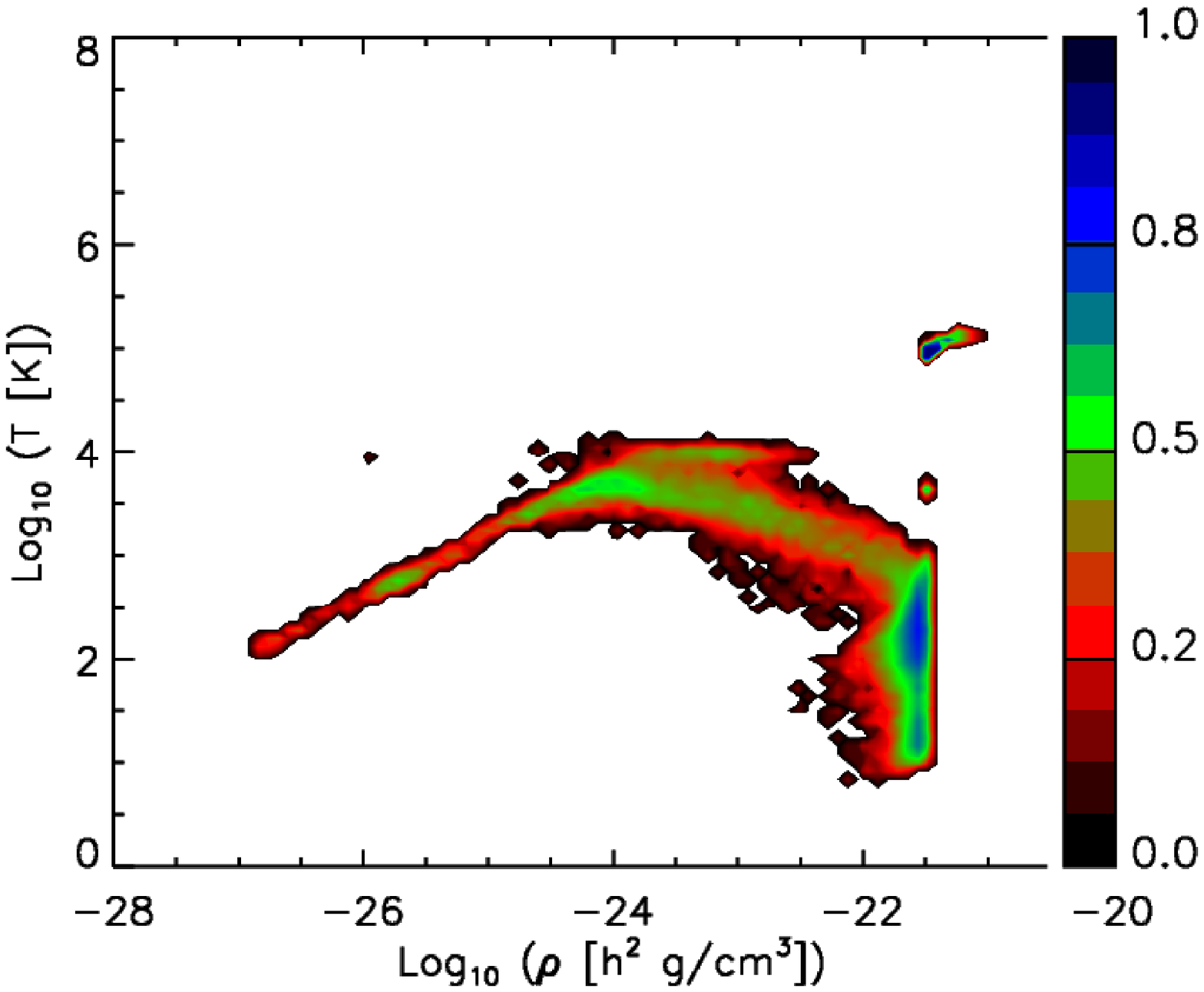}\\
\flushleft{$z=15.00$}\\
\caption[Probabilities at $z=15$]{\small
Phase distributions for metal enriched particles at redshift $z=18.01$ (upper row), $z=15.00$ (central row), and $z=9.00$ (lower row), for \fnl=0 (left column), \fnl=100 (central column), \fnl=1000 (right column).
The x-axes are comoving densities, and the y-axes are temperatures, in logarithmic scale; the colors refer to the probability distributions of the enriched particles.
}
\label{fig:phase}
\end{figure*}

In  Fig.~\ref{fig:phase}, we show phase diagrams at different redshift for the enriched particles, color-coded according to their probability distribution.
Early star formation episodes, dominated by massive, short-lived population III stars \cite[][]{Maio2010} quickly enrich the surrounding medium, from the higher to the lower densities.
The timing of these episodes is affected by the adopted \fnl, though, mostly at very high redshift.
Differences for \fnl=0 and \fnl=100 are quite small, but they become much more evident for \fnl=1000.
In this latter case, in fact, the densities (see also next sections) are strongly skewned towards higher values, thus gas cooling, condensations, and fragmentation is more enhanced and lead to an earlier onset of star formation \cite[Fig. 4 in][]{MaioIannuzzi2011,Maio2011cqg}.
Indeed, first star formation events are already detected at $z\sim 23$ (when the cosmic time is about $144$~Myr) for the \fnl=1000 case, while they appear only at $z\sim 19-20$ ($\sim 45-32$~Myr later) for \fnl=0-100.
Given the rapid evolution of primordial massive stars ($\lesssim 20$~Myr), metal pollution in the \fnl=1000 case has already taken place, when the first stars are formed in the models with \fnl=0 and \fnl=100.
By comparing the time sequence of the pollution patterns for \fnl=0, \fnl=100, and \fnl=1000, we note that at $z\sim 18$ they show quite different stages.
For  \fnl=0 and \fnl=100, the first metals are just being ejected from newly born stars in high-density environments (and the first heated particles are also visible at $T\sim 10^{5}\,\rm K$), while for \fnl=1000 the enrichment process is in a more advanced stage, with enriched materials spread much further out, and reaching also very underdense regions.
\\
However, the following spreading episodes mitigate the effects of non-Gaussianities and by $z\sim 15$ (i.e. in roughly $60$~Myr) the metal distributions in the phase space become quite similar.

\subsubsection{Probability distributions}

\begin{figure*}
\centering
\flushleft{\fbox{\fnl=0}}
\includegraphics[width=\textwidth]{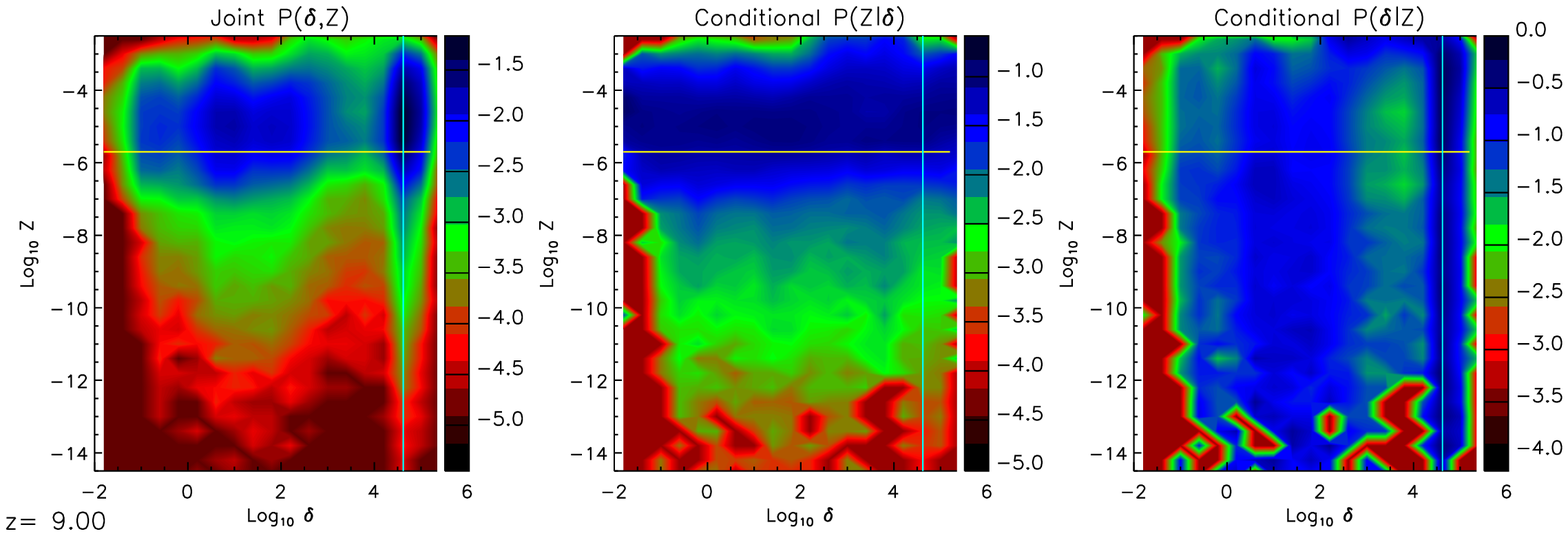}\\
\flushleft{\fbox{\fnl=100}}
\includegraphics[width=\textwidth]{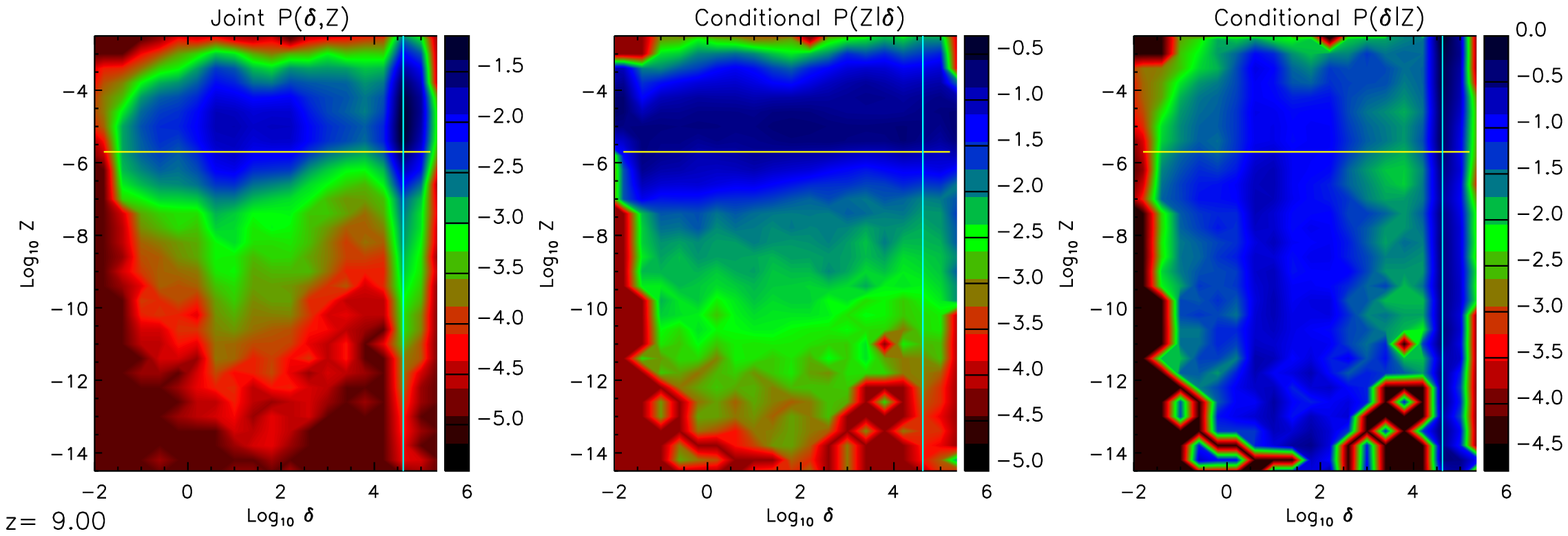}\\
\flushleft{\fbox{\fnl=1000}}
\includegraphics[width=\textwidth]{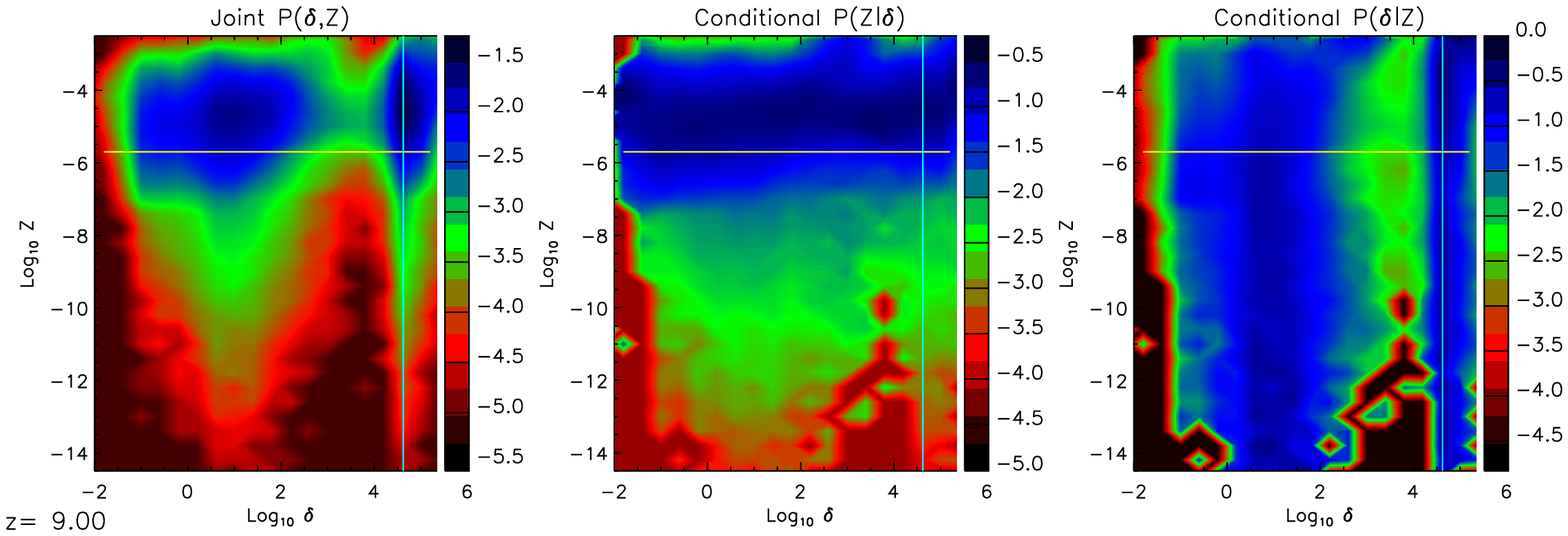}\\
\caption[Probabilities at $z=9$]{\small
Joint probability distribution $P(Z,\delta)$ (left column) and the conditional probability distributions $P(Z|\delta)$ (central column) and $P(\delta|Z)$ (right column), for \fnl=0 (upper row), \fnl=100 (central row), and \fnl=1000 (right row), at redshift $z=9.00$.
The vertical lines refer to the density threshold for star formation \cite[for details][]{MaioIannuzzi2011}, while the horizontal line correspond to the critical metallicity, $Z_{crit}=10^{-4}Z_\odot$.
}
\label{fig:P9}
\end{figure*}

The trends for metal spreading are supported by the probability distribution functions displayed, for sake of completeness, in Fig.~\ref{fig:P9}.
For these calculations we define the overdensity variable:
\begin{equation}
\label{delta}
\delta\equiv \rho/\rho_{cr},
\end{equation}
with $\rho$ gas density,
\begin{equation}
\label{rho_cr}
\rho_{cr} = \frac{3H^2}{8\pi G},
\end{equation}
the expansion parameter
\begin{equation}
\label{H}
H = H_0 \left[ (1-\Omegat)(1+z)^2 + \Omegam(1+z)^3 + \Omegal \right]^{1/2},
\end{equation}
and $\Omegat = \Omegam +\Omegal$
\cite[][]{Peebles1993,Peacock1999,ColesLucchin2002}.
\\
In Fig.~\ref{fig:P9}, we consider metallicity, $Z$, and overdensity, $\delta$ -- as defined in the previous eq. (\ref{delta}), (\ref{rho_cr}), and (\ref{H}).
The plots show the joint probability distributions $P(Z,\delta)$ (left column) and the conditional probability distributions $P(Z|\delta)$ (central column) and $P(\delta|Z)$ (right column), for \fnl=0 (upper row), \fnl=100 (central row), and \fnl=1000 (lower row), at redshift $z=9.00$.
In the plots it is evident the efficient enrichment process spanning a large range of over-densities and metallicities.
Metals are produced at high densities ($\delta \gtrsim 10^4$) and ejected by wind feedback into the surrounding pristine regions.
The high metal yields allow most of the involved gas to be rapidly polluted to metallicities $Z\gtrsim Z_{crit}$ and to pass to the popII-I regime, almost independently of \fnl, as well demonstrated by all the probability distributions.
The differences surviving at this redshift are difficult to detect.
The joint probability distributions (left column) for \fnl=0 and \fnl=100 are basically identical, with most of the gas at $10^{-6}<Z<10^{-4}$ and some residual gas below $\sim 10^{-6}$.
Also the \fnl=1000 case is very close to the previous ones, and no significant differences are found.
\\
The conditional probability distributions (central and right columns) show some slight deviations of the \fnl=1000 case with respect to \fnl=0 and \fnl=100, but statistically they are not relevant, and all the cases seem to have gas enriched well above $Z_{crit}$ and spread down to $\delta\sim 10^{-1}-10^{-2}$.
\\
Comparing to the findings regarding higher redshift (previous section), we can state that differences in the enrichment episodes for different non-Gaussian cosmologies are occurring mainly in primordial epochs, and therefore, the resulting metal filling factor could have signatures depending on the various scenarios.


\subsubsection{Filling factors}
\begin{figure}
\centering
\includegraphics[width=0.42\textwidth]{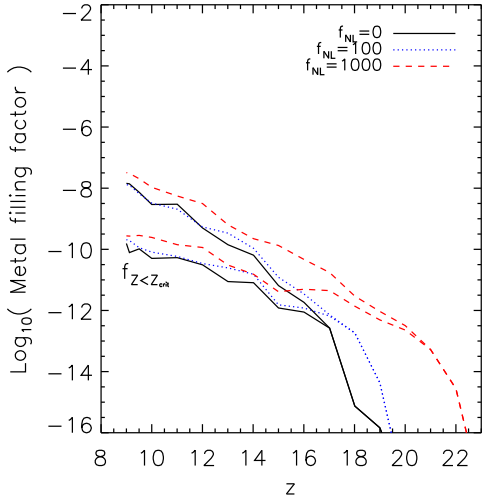}
\caption[Metal filling factor evolution]{\small
Metal filling factors as function of redshift in the 0.5~\Mpch{} side boxes (left), and 0.5~\Mpch{} side box (right), for \fnl=0 (solid lines), \fnl=100 (dotted lines), and \fnl=1000 (dashed lines). Individual contributions from popIII regions ($0<Z<Z_{crit}$, bottom lines) are also shown, as indicated by the legends).
}
\label{fig:ffcompare}
\end{figure}

As very first enrichment episodes can have distinct impacts in Gaussian and non-Gaussian models, it is relevant to quantify the metal filling factor at early times.\\
In  Fig.~\ref{fig:ffcompare}, we show the redshift evolution of the metal filling factors, $f_V$, for the three cases considered, \fnl=0, \fnl=100, \fnl=1000.
We define $f_V$ according to:
\begin{equation}
f_V \equiv \frac{\sum_i m_{Z,i}/\rho_{Z,i}}{\sum_j m_j/ \rho_j} \sim \frac{\sum_i m_{Z,i}/\rho_{Z,i}}{V},
\end{equation}
with $i$ and $j$ integers running over the number of the enriched particles and of all the SPH particles, respectively,
$m_{Z}$ particle metal mass,
$m$ particle total mass,
$\rho_{Z}$ metal mass density,
$\rho$ total-mass density,
and $V$ simulation volume.
We perform the calculation for the whole simulations and, additionally, by distinguishing for the two stellar populations (popIII and popII-I).
In Fig.~\ref{fig:ffcompare}, we plot the evolution of the global filling factors and of the popIII filling factors (lower lines, denoted by $f_{Z<Z_{crit}}$).
The redshift interval (from $z\sim 23$ to $z\sim 9$) covers a period of $\sim 400$~Myr, from when the Universe was $\sim 140$~Myr old, to when it was $\sim 540$ Myr old.
The results show well the distinct behaviours of the different models at early times, with an enrichment process which is earlier in the \fnl=1000 case and later in the \fnl=0 case.
At redshift $z\sim 19$, the pollution levels differ of about two orders of magnitude between \fnl=0 and \fnl=100, and between \fnl=100 and \fnl=1000.
The continuous spreading events alleviate the delay between \fnl=0 and \fnl=100 to a factor of a few by $z\sim 16$, and lead to full convergence by $z\sim 12$.
Also the advanced status of the enrichment process in the \fnl=1000 model is gradually reduced to about one order of magnitude by $z\sim 14$ and achieves almost convergence at $z\sim 9$.
The trends for the popIII regimes are similar and their contribution to the total filling factor always drops down to $\sim 1\%$ between $z\sim 16$ and $z\sim 9$ (i.e. in $\sim 300$~Myr), despite the different \fnl{} values adopted.
The transition from the primordial popIII regime to the following popII-I regime is regulated by the stellar lifetimes and initial mass function of the first populations \cite[][]{Maio2010}, so different assumptions on the popIII IMF would change the relative contributions of the popIII regime and the onset of metal pollution, but would not substantially modify the conclusions on the effects on non-Gaussianities.
Indeed, the different behaviours are consequences of the earlier onset of star formation in the higher-\fnl{} model \cite[][]{MaioIannuzzi2011,Maio2011cqg}.

\subsection{Implications for cosmic haloes}\label{Sect:haloes}
For a further investigation, we study the implications of feedback mechanisms on the cosmic haloes in Gaussian and non-Gaussian models.
As mentioned earlier, the haloes and their respective dark, gaseous, and stellar components are found by use of a FoF algorithm.


\subsubsection{Cosmic gas within formed structures}
\begin{figure}
\centering
\includegraphics[width=0.4\textwidth]{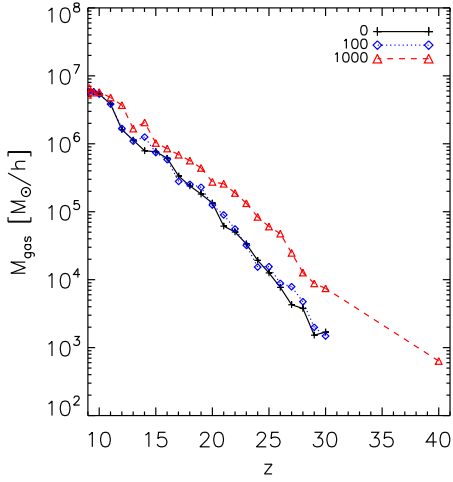}
\caption[Mass evolution]{\small
Mass (top) and fraction (bottom) evolution of the gaseous components for the biggest halo as a function of redshift, for \fnl=0 (solid lines, cross symbols), \fnl=100 (dotted lines, rhombic symbols), and \fnl = 1000 (dashed lines, triangular symbols).
}
\label{fig:mass}
\end{figure}

In Fig.~\ref{fig:mass} we plot the redshift evolution of the gas mass in the largest objects and the corresponding gas fraction\footnote{ We note that at high redshift the number of particles per halo is rather small, and so the error on the estimated quantities is large.}
.
Also in this case, for \fnl=0 and \fnl=100 there are no evident differences, while for \fnl=1000 there is an earlier formation of the structures, consistently with the previous sections.
\\
The mass growth is quite different at very early epochs and in fact, in the \fnl=0 and \fnl=100 cases $\sim 10^3\,\msunh$ objects are formed at $z\sim 30$, while in the \fnl=1000 case they are formed already at $z\sim 40$, about 40~Myr earlier, and are almost one order of magnitude larger at $z\sim 30$.
Such huge discrepancy shrinks over the time and is reduced down to a factor of $\sim 2$ at $z\sim 20$, when the gas masses contained in the primordial haloes are $\sim 10^5\,\msunh$, and completely disappears at redshift $z\lesssim 15$.
\\
We note the gas that gets accumulated within early structures has a direct connection to the growth of the dark-matter haloes and thus reflect the original differences in \fnl.
The baryonic build-up, with the catch-up of the dark-matter structures, is well visible at high redshift, when gas is falling into the first dark-matter haloes and starts cooling.
The first SN bursts and feedback effects are then the main responsible for gas expulsion from star formation events, and for spreading material into the low-density environments by winds.
\\
These processes contribute to settle down a self-regulated regime in which infalling gas is gradually converted into stars and feedback mechanisms control the amount of material which is expelled.

\subsubsection{Halo profiles}
\begin{figure}
\centering
\includegraphics[width=0.45\textwidth]{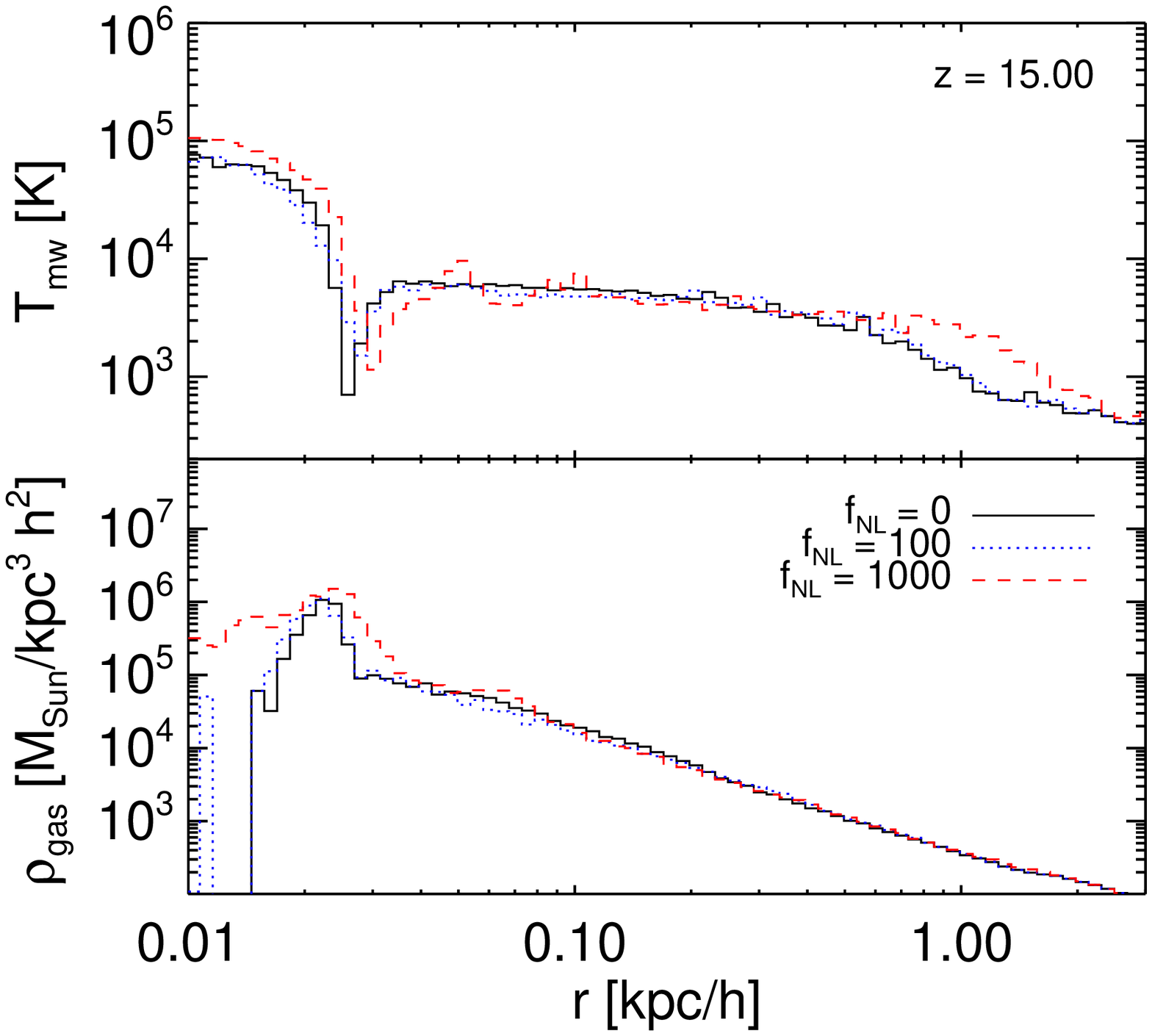}
\includegraphics[width=0.45\textwidth]{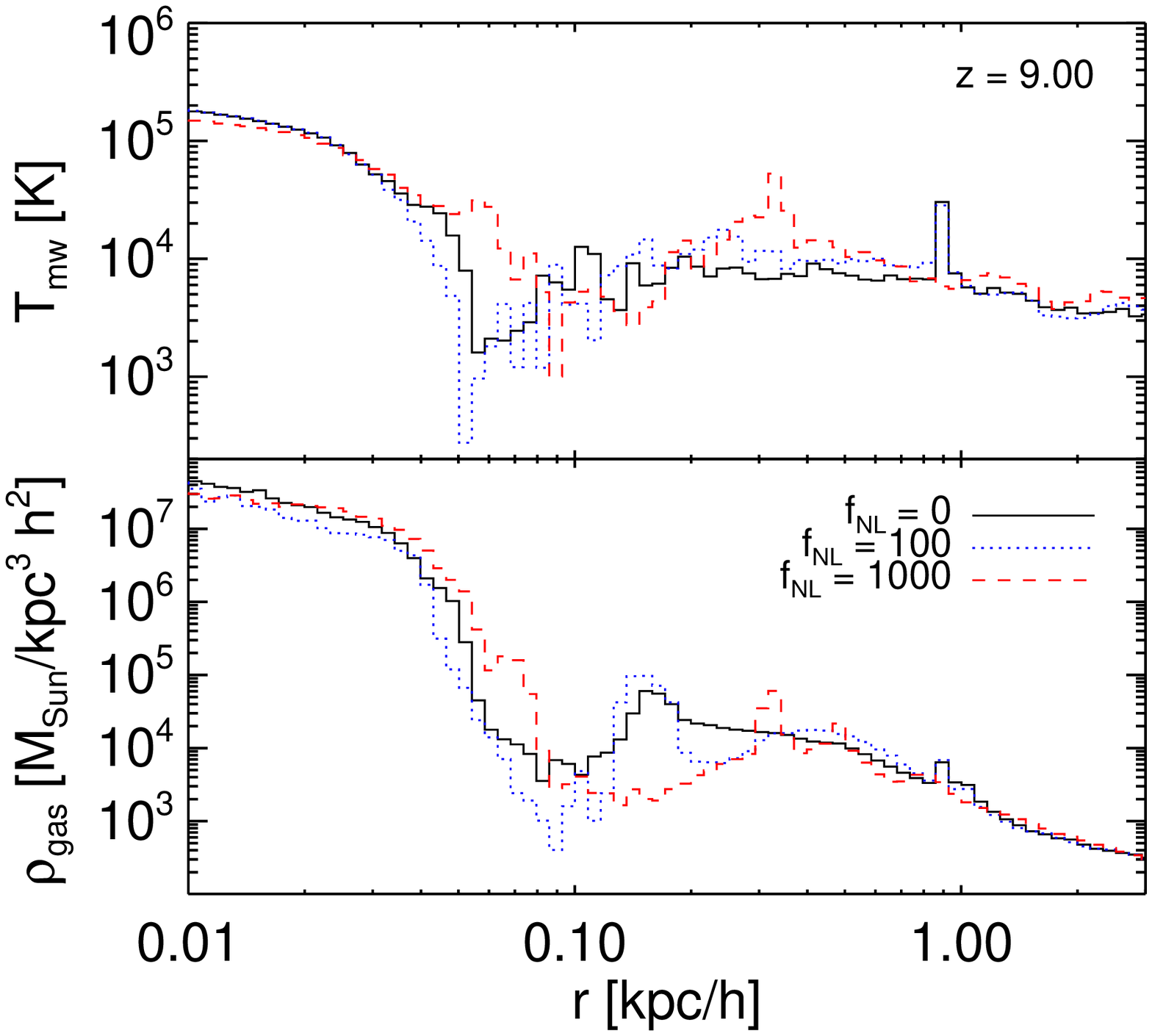}
\caption[Profiles]{\small
Mass-weighted temperature, $T_{mw}$ (upper panels), and mass density, $\rho_{gas}$ (lower panels), profiles as a function of physical radius, $r$, at redshift $z=15$ (top) and $z=9$ (bottom), for \fnl=0 (solid lines), \fnl=100 (dotted lines), and \fnl=1000 (dashed lines).
}
\label{fig:profiles}
\end{figure}

To conclude, in Fig.~\ref{fig:profiles}, we study the effects on the profiles of the largest haloes at different epochs.
\\
Broadly speaking, the temperature profiles (upper panels) are characterized by high central values ($\sim 10^4-10^5\,\rm K$), where most of the star formation is taking place, and lower ones in the peripheral regions, where the gas is being shock-heated to $\sim 10^3-10^4\,\rm K$ and subsequently cools down via molecular emissions to some hundreds Kelvin (e.g. the drops at $\sim 20-30$~pc/$h$ at $z=15$, or at $\sim 50-100$~pc/$h$ at $z=9$).
At redshift $z=15$, during the first stages of star formation, non-Gaussianities are slightly visible for \fnl=1000 and the only effect is a larger temperature of a factor of $\sim 2-3$ at a distance of $\sim 1\,\kpch$ from the center.
This is related to the larger halo mass, which results in higher shock heating temperatures, and in an indirect effect of \fnl.
At redshift $z=9$, star formation and feedback mechanisms have been going on for almost half a Gyr, the turbulent state of the medium heavily dominates the profiles, and it becomes impossible to identify trends correlated to \fnl.
\\
Density profiles (lower panels) have a very regular, smooth shape at $z=15$, and the most clumped regions correspond to the coldest temperatures.
At this epoch, the larger concentrations in the core of the structures could be associated to non-Gaussianities, mostly for \fnl=1000.
Precisely, this is a more directly linked to the earlier build-up of halo masses in the \fnl=1000 model.
Later on, the cumulative development and propagation of shocks from SN explosions determine a complex interplay between infalling cold material and expanding hot gas \cite[see a detailed treatment in][]{Maio2011}, so, the continuous gas compressions and rarefactions (as e.g. at radii of $\sim 100$~pc/$h$) erase the signal from different \fnl{} and wash it out by $z\sim 9$.

\subsection{Feedback, non-Gaussianities and primordial bulk flows}\label{Sect:vbulk}
\begin{figure}
\centering
\includegraphics[width=0.42\textwidth,height=0.3\textheight]{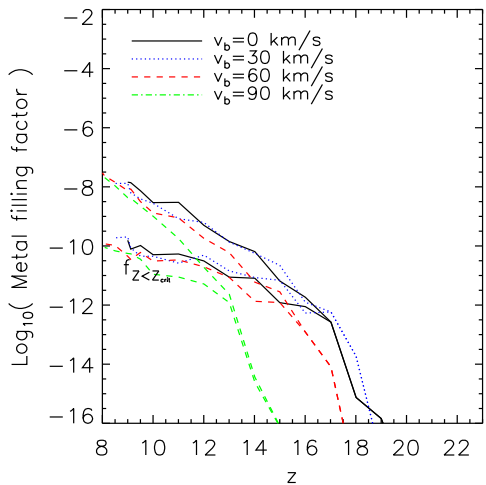}
\caption[Metal filling factor evolution and primordial bulk flows]{\small
Metal filling factors as function of redshift for \fnl=0 and primordial gas bulk velocity \vb=0 km/s (solid lines), \vb=30 km/s (dotted lines), \vb=60 km/s (dashed lines), and \vb=90 km/s (dot-dashed lines). Individual contributions from popIII regions ($0<Z<Z_{crit}$, bottom lines) are also shown, as indicated by the label.
}
\label{fig:ffvb}
\end{figure}

We conclude by briefly showing the effects of primordial streaming motions on the formation of primordial structures and metal enrichment from first stars, in Gaussian and non-Gaussian models.
As an example, in Fig.~\ref{fig:ffvb}, we plot the metal filling factor for the high resolution simulations with \fnl=0 re-run with initial primordial gas streaming motions at decoupling of \vb=0, 30, 60, 90 km/s.
Such primordial steaming motions determine gas bulk flows on $\sim$~Mpc scales and hinder gas condensation in mini-haloes with masses below $\sim 10^{8}\,\msunh$ \cite[][]{TseliakhovichHirata2010,Maio2011}, because larger potential wells are needed to equal the additional contribution to gas kinetic energy.
As a result, the onset of star formation and consequent feedback mechanisms are delayed to lower redshift, when the haloes have grown enough to trap the gas and let it cool and condense.

\section{Discussion and conclusions}\label{Sect:discussion}
In this work, we have discussed the principal implications of feedback effects on non-Gaussianities, by using the N-body/SPH chemistry simulations, presented by \cite{MaioIannuzzi2011}.
\\
Besides gravity and hydrodynamics \cite[][]{Springel2005}, the runs include radiative gas cooling both from molecules and atomic transitions \cite[according to][]{Maio2007}, multi-phase model \cite[][]{Springel2003} for star formation, UV background \cite[][]{HaardtMadau1996}, wind feedback \cite[][]{Springel2003,Aguirre_et_al_2001}, chemical network for e$^-$, H, H$^+$, H$^-$, He, He$^+$, He$^{++}$, H$_2$, H$_2^+$, D, D$^+$, HD, HeH$^+$ \cite[e.g.][and references therein]{Yoshida2003,Maio2006,Maio2007}, and metal pollution from popIII and/or popII-I stellar generations, ruled by a critical metallicity threshold of $Z_{crit}=10^{-4}\,\zsun$ \cite[][]{Tornatore2007,Maio2010,Maio2011b,Maio2011cqg}.
Local non-Gaussianities are included by means of the \fnl{} parameter, here chosen to be \fnl=0, \fnl=100, or \fnl=1000, in order to explore thermodynamical and statistical properties of gas evolution and of the impacts of feedback effects on cosmic structures in different non-Gaussian universes.\\
Since non-Gaussianities are based on the primordial matter density field, it is natural to start our study from the gas cumulative distribution.
The different models show deviations at large densities, consistently with the biased initial conditions in larger-\fnl{} models, and at high $z$, first-molecule formation and catastrophic gas collapse are strongly enhanced by larger non-Gaussian parameters.
The global thermodynamical evolution of the gas 
shows accordingly significant deviations from the Gaussian case mostly at high densities, for larger \fnl.
As a consequence, star formation and heating due to PISN/SN explosions take place earlier and chemical feedback, as well.
Metal spreading 
preserve some information on the original \fnl{} values with wider areas involved by the pollution process for \fnl=1000 and smaller areas for \fnl=100 and \fnl=0.
The differences are more visible during primordial epochs ($z\gtrsim 15$) and less and less at later times ($z \lesssim 9$), as well demonstrated by phase 
and probability distributions. 
The enrichment process is very efficient in polluting the medium and even underdense regions ($\delta < 1$), achieving $Z\gtrsim Z_{crit}$ in few tens of Myr.
The metal filling factor 
initially (at $z\gtrsim 15$) shows discrepancies up to $\sim 2$ orders of magnitude, but successively the trends rapidly converge within a factor of a few (at redshift $z\sim 9$), for both popIII and popII-I regions.
By looking at the largest-halo mass evolution, 
it emerges that not only early structures form earlier for larger \fnl, but also that their potential well growth can trap more gas, because of the larger halo masses.
At lower redshift, the trends converge, regulated by star formation processes.
The most-massive halo temperature and density profiles 
confirm that and suggest the fundamental reason why signatures of non-Gaussianities in the luminous structures are washed out: i.e., mechanical and chemical feedback strongly perturb the surrounding medium, the gas behaviour becomes soon dominated by shocks and turbulence and, thus, looses memory of its original distribution.
\\
We stress that, despite the small box size (which prevents extremely large objects to be formed), the high resolution ($\sim 40\,\msunh$ per gas particle) of the simulations enabled us to check for the first time the effects of non-Gaussianities on quite tiny scales (up to a few parsecs), at high redshift.
In general, there is a delay of $\sim 50$~Myr in \fnl=0 and \fnl=100, with respect to \fnl=1000.
These effects are visible at very early times, without the need for implementing further non-Gaussian corrections \cite[like the ones related to \gnl,\ or \tnl{} parameters, as also discussed in detail by][]{MaioIannuzzi2011,Maio2011cqg}.
\\
Whereas, the inclusion of non-linear effects that imply primordial gas streaming motions with velocities \vb$\sim 30\, \rm km/s$ \cite[e.g.][]{TseliakhovichHirata2010,Maio2011} could slightly alter the initial phases of star formation \cite[as demonstrated by][]{Maio2011} and add some degeneracies with the \fnl{} effects \cite[see][]{MaioIannuzzi2011}.
Substantial differences on the feedback impacts are not expected, though. 
\\
Some changes would appear in case of a different popIII IMF, which, at the present, is basically unknown, and would be mainly related to the different lifetimes of the different stellar masses sampled \cite[see also][]{MaioIannuzzi2011}.
For the top-heavy popIII IMF considered here, the mass range was between 100$\msun$ and 500$\msun$, and the first stars to die and pollute the medium were the PISN, in the range [140, 260]~$\msun$, after at most 2~Myr.
For low-mass popIII IMF the first stars exploding would be $\sim 10 \msun$ SNe, after $\sim 10-100~$Myr.
This means that the stellar evolution process of low-mass ($\lesssim 100\msun$) stars would slow down the activation of feedback mechanisms and postpone the spreading events of some tens or a hundred Myr.
\\
In general, the main differences among the various cosmologies are detected at high redshift, when luminous structures start their life.
This happens because larger \fnl{} parameters determine a density field which is biased to higher densities, lead to larger most massive haloes, and, as a consequence, imply shorter cooling times for the amount of gas that accumulates in the center.
Since stars form earlier, metal enrichment takes place earlier, too.
\\
Thus, all these processes can be seen as derived effects of the non-Gaussian initial conditions.
In particular, the essential point is the the formation time-scales of high-$z$ structures, that influence the initial evolution of the first objects, accordingly to the \fnl{} values considered.
On the other hand, it seems that feedback mechanisms are the key ingredient which largely wash out such original discrepancies and hinder the possibility of distinguishing the different models at later times (e.g. at $z \lesssim 9$).
They, indeed, affect the amount of available cold star forming medium, and regulate the star formation activity, independently from the initial conditions.
That is the reason why all the trends converge and are not tightly linked to the non-Gaussian parameters.
\\
From our discussion it emerges that studies and observations of the primordial Universe can be very useful tools to try to address non-Gaussianities, much more powerful than low-redshift investigations.
In fact, signatures from first star formation episodes could be potential probes of primordial non-Gaussianities during the first Gyr of the Universe.
For example, high-redshift, long gamma-ray bursts \cite[][]{Cucchiara2011} seem to be highly correlated with popIII star formation, as recently demonstrated by e.g. \cite{Campisi2011}.
So, they are suitable probes of the conditions of the Universe at early times, when non-Gaussian effects could be still visible in the baryon evolution.
Similarly, first stars and galaxies are detected up to $z\sim 8-10$ \cite[][]{Bouwens2011} and future instruments (like JWST) will likely see more and more of these objects, providing important information about the status of the Universe in its early infancy and possibly about the cosmic matter distribution.
Primordial episodes of star formation and metal enrichment from high-redshift objects have a very steep evolution (see Fig.~\ref{fig:ffcompare}), and, even if the first onsets differ by only $\Delta z \simeq 0.5$ for \fnl=0 and \fnl=100, the corresponding metal filling factors differ by $\sim 2$ orders of magnitudes.
Despite the current difficulty of observing metal filling factors at such high redshift, they might provide an interesting opportunity for future observational missions.



\bibliographystyle{mn2e}
\bibliography{bibl.bib}

\label{lastpage}
\end{document}